\documentclass[fleqn,usenatbib]{mnras}

\usepackage[T1]{fontenc}
\DeclareRobustCommand{\VAN}[3]{#2}
\let\VANthebibliography\thebibliography
\def\thebibliography{\DeclareRobustCommand{\VAN}[3]{##3}\VANthebibliography}

\usepackage{graphicx}	
\usepackage{amsmath}	
\usepackage{amssymb}	
\usepackage[flushleft]{threeparttable}

\newcommand{\mdot}{$\dot{m}_{\rm Edd}$}	
\newcommand{\rg}{$R_{\rm g}$}	

\title[Time-lags in AGN]{Modeling the UV/optical continuum time-lags in AGN}

\author[E. S. Kammoun et al.]{
E. S. Kammoun,$^{1,2}$\thanks{E-mail: \url{ekammoun@irap.omp.eu}}
I. E. Papadakis,$^{3,4}$
M. Dov{\v c}iak$^{5}$
\\
$^{1}$Department of Astronomy, University of Michigan, 1085 South University Avenue, Ann Arbor, MI 48109-1107, USA\\
$^{2}$ IRAP, Universit\'{e} de Toulouse, CNRS, UPS, CNES, 9, Avenue du Colonel Roche, BP 44346, F-31028, Toulouse Cedex 4, France \\
$^{3}$Department of Physics and Institute of Theoretical and Computational Physics, University of Crete, 71003 Heraklion, Greece\\
$^{4}$Institute of Astrophysics, FORTH, GR-71110 Heraklion, Greece\\
$^{5}$Astronomical Institute of the Czech Academy of Sciences, Bo{\v c}n{\'i} II 1401, CZ-14100 Prague, Czech Republic
}

\date{Accepted XXX. Received YYY; in original form ZZZ}

\pubyear{2021}

\begin{document}
\label{firstpage}
\pagerange{\pageref{firstpage}--\pageref{lastpage}}
\maketitle

\begin{abstract}
Thermal reverberation in accretion discs of active galactic nuclei is thought to be the reason of the continuum UV/optical time lags seen in these sources. Recently, we studied thermal reverberation of a standard Novikov-Thorne accretion disc illuminated by an X--ray point-like source, and we derived an analytic prescription for the time lags as function of wavelength. In this work, we use this analytic function to fit the time-lags spectra of seven Seyferts, that have been intensively monitored, in many wave-bands, in the last few years. We find that thermal reverberation can explain the observed UV/optical time lags in all these sources. Contrary to previous claims, the magnitude of the observed UV/optical time-lags is exactly as expected in the case of a standard accretion disc in the lamp-post geometry, given the black hole mass and the accretion rate estimates for the objects we study. We derive estimates of the disc accretion rates and corona height for a non-spinning and a maximally spinning black hole scenarios. We also find that the modelling of the continuum optical/UV time-lags can be used to estimate the black hole spin, when combined with additional information. We also find that the model under-predicts the observed X--ray to UV time-lags, but this difference is probably due to the broad X-ray auto-correlation function of these sources. 

\end{abstract}



\section{Introduction}
\label{sec:intro}

Accretion of matter into a supermassive black hole (BH) is currently believed to be responsible for the enormous power emitted by active galactic nuclei (AGN). Disc reverberation mapping \citep[][]{Blandford82, Peterson14} is one of the best ways to test  the geometry and the accretion  flow  in these objects. This technique utilizes time-lags between the continuum variations at different wavelengths\footnote{Reverberation mapping also uses time-lags between continuum and emission lines like,  e.g. H-beta reverberation in the optical band and Fe K lines in the X-ray band}. Recently, in \cite{Kammoun19lag} and \citet[hereafter, K20]{Kammoun20}, we studied the UV/optical time-lags as a function of wavelength (i.e., the so-called ``time-lags spectrum''), assuming an  X-ray  point-like  source  illuminating a standard Novikov-Thorne accretion disc \citep{Novikov73}, i.e., assuming the so-called `lamp-post geometry' \citep[e.g.,][]{George89,Matt1991}. In K20, we considered in detail all special and general relativity effects, and the disc ionization when computing the disc reflection flux. We investigated the dependence of the time-lags on BH mass and spin, X-ray corona height,  luminosity  and photon index, as well as disc accretion rate,  inclination, and inner/outer radii. We also presented an analytic function for the time-lags spectra, which can be used to fit the observed time-lags spectra. 

Intensive, X--ray to UV/optical campaigns of a few bright AGN have been performed the last few years. One of the main results from these campaigns is that the UV/optical variations are well correlated, but with a delay, which propagates from the UV to the optical bands, as expected in the case of X--ray illumination of the disc. One of the main results from these campaigns is that the dependence of the time-lags on wavelength agrees with the predictions in the case of a standard, thin accretion disc, but the lags amplitude is larger than expected, given the BH mass and accretion rate of the objects, by a factor of $\sim 2-3$. 

In this work, we present the results from a detailed modelling of the observed time-lags spectra, using the K20 model. We focus on the modelling of the UV/optical time-lags only. In Section~\ref{sec:sample}, we present the sample analysed in this work. In Section~\ref{sec:Xrayspec}, we present the X-ray spectral fits. A comparison between the K20 and time-lags spectra presented in previous works, and the time-lag fits are presented in Section~\ref{sec:lagfit}. We finally discuss the results and present our conclusions in Section~\ref{sec:discussion}.

\begin{table}
\caption{The sources in our sample. Luminosity distance, $D_{\rm L}$, BH mass, $M_{\rm BH}$, accretion rate in Eddington units, \mdot,  and the time range we used to extract the \textit{Swift}/XRT spectra are listed in columns 3-6 (see text for details).}
\label{tab:sources}
\resizebox{\columnwidth}{!}
{
\begin{threeparttable}

\begin{tabular}{llllll} 
\hline
Source	    &	Ref.        & $D_{\rm L}$	& $M_{\rm BH}^{\dagger}$    &	\mdot	&	Time range	        	\\
	        &	            & (Mpc)	        & $	  (10^7~M_\odot$)       &		    &	(MJD)	                \\ \hline
Mrk~142 	& C20	        & 199.1	        & $0.197_{-0.04}^{+0.04}$   &	0.86	 &  $58526.5-58545.5	$    	\\ [0.1cm]
Mrk~509	    & E19	        & 151.2	        & $11.19_{-0.86}^{+0.94}$   &	0.14$^\ast$	&  $57974-58020$  \\ [0.1cm]
NGC~2617	& F18 	        & 61.5	        & $3.24_{-2.14}^{+6.31}	$   &	0.01	&  $56684-56726$  \\ [0.1cm]
NGC~4151	& E17 	        & 16.6      	& $	3.59^{+0.45}_{-0.38}$   &	0.02$^\ast$	&  $57486-57493.5$ \\ [0.1cm]
NGC~4593	& C18 	        & 38.8	        & $ 0.76^{+ 0.16}_{-0.16}$  &	0.022	&  $57588.5-57593.5	$  \\ [0.1cm]
NGC~5548	& F16 	        & 74.5	        & $	5.23 \pm0.19$           &	0.05	&  $	- $             \\ [0.1cm]
NGC~7469	& P20 	        & 70.8	        & $	0.90 \pm 0.1$           &	0.3	    &  $56460-56480$  \\ \hline

\end{tabular}
\begin{tablenotes}
\item[$^\dagger$]$M_{\rm BH}$ are from ``The AGN Black Hole Mass Database'' \citep{Bentz15}; we use the estimates when considering all emission lines. $M_{\rm BH}$  for NGC~2617 is taken from \cite{Fausnaugh17}. The light curves and time-lags are taken from: C20 \citep{Cackett20}, E19 \citep{Edelson19}, F18 \citep{Fausnaugh18}, E17 \citep{Edelson17}, C18 \citep{Cackett18}, F16 \citep{Fausnaugh16}, and P20 \citep{Pahari20}. We computed $D_{\rm L}$ using the source redshift,  assuming a flat Universe with: $H_0 = 70~\rm km~s^{-1}~Mpc^{-1}$, $\Omega_{\rm \Lambda} = 0.7$, and $\Omega_{\rm M} = 0.3$. $D_{\rm L}$ for NGC~4151 was taken from \cite{Bentz2013}. 
\item[$\ast$] Accretion rates are taken from references listed in the second column, except for Mrk~509 and NGC~4151, which are taken from \cite{Lubinski16}. 
\end{tablenotes}
\end{threeparttable}
}
\end{table}

\section{The sample}
\label{sec:sample}

We selected AGN with known time-lag spectra, based on intensive, multi-wavelength campaigns. Our search resulted in nine objects: Mrk~142, Mrk~509, NGC~2617, NGC~4151, NGC~4593, NGC~5548, NGC~7469, MCG$+$08-11-011, and Fairall~9 . We exclude the latter two sources from the current analysis. The MCG$+$08-11-011 data \citep{Fausnaugh18} lack simultaneous X-ray observations, which are essential for our modeling. As for Fairall~9, modelling its time-lag spectrum \citep{Santisteban20} will be left for a dedicated paper, in preparation. 

Thus, we are left with seven sources with reported UV/optical time-lags and simultaneous \textit{Swift}/XRT observations. They are  listed in Table~\ref{tab:sources}, together with references for the time-lags spectra we used, luminosity distances, BH mass and accretion rate estimates (second, third, fourth and fifth column, respectively). We note that data for two of the sources (NGC~4593 and NGC~4151) were presented in various papers. \cite{Mchardy18} presented the results from the analysis of the \textit{Swift} monitoring data of NGC~4593. Later on, \cite{Cackett18} presented \textit{HST} monitoring observations and re-analysed the \textit{Swift} data. The results of these two papers are in agreement and, for consistency, we adopted the time-lag values from \cite{Cackett18}. As for NGC~4151, we use the time lags presented by \cite{Edelson17}, that are in agreement with the \cite{Edelson19} results. Accretion rates (in Eddington units) are listed in the fifth column of Table~\ref{tab:sources}. The Mrk~142, NGC~2617, NGC~4593, and NGC~5548 are taken from \cite{Cackett20}, \cite{Fausnaugh18}, \cite{Cackett18}, and \cite{Fausnaugh16}, respectively, and are based on bolometric luminosity estimates using the average, 5100~\AA\ AGN flux. The NGC~4769 estimate is taken from \cite{Pahari20}, and is based on a bolometric luminosity estimation, using the X--ray luminosity. Finally, the \mdot\ estimates for Mrk~509 and NGC~4151 are taken from \citep{Lubinski16}, and are based on X-ray/UV/optical model fits. We note that for all sources we used the centroid values of the ICCF as reported in the corresponding papers.

\section{X-ray spectral fits}
\label{sec:Xrayspec}

As discussed in K20, proper modelling of thermal reverberation in AGN requires knowledge of the X-ray spectral shape and luminosity. This is available for NGC~5548 \citep{Mathur17}. We therefore adopted their best-fit values of $\Gamma = 1.5$ and $L_{\rm X, Edd} = 0.0034$ (where $L_{\rm X, Edd}$ is the observed, $2-10$~keV X--ray luminosity, in Eddington units). For the rest of the sources, we first considered the $2-10$~keV \textit{Swift}/XRT light curves, and we defined time intervals where the count rate is close to the mean count rate (listed in the last column of Table~\ref{tab:sources}). We then extracted the X--ray spectra from these intervals using the automatic \textit{Swift}/XRT data products generator\footnote{\url{https://www.swift.ac.uk/user_objects/}} \citep{Evans09}.

\begin{table*}
	\caption{Best-fit parameters obtained by fitting the X-ray spectra of the sources analysed in this work.}
	\label{tab:spectra_fits}
	\begin{threeparttable}
\begin{tabular}{lclllclcll} 
\hline
Source	&$	\log F_{2-10}^{a}	$&	$L_{\rm X, Edd}$	&$	\Gamma	$&$	N_{\rm H}^b	$&$	\log \xi_{\rm abs}^{c}	$&$	f_{\rm cov}	$&$	\log \xi_{\rm refl}^{c}	$&$	\rm Norm_{\rm refl}	$&	$\rm \chi^2/dof$	\\ \hline
Mrk~142	&$	-11.89 $ &	0.0246	&$	 2.37 \pm 0.04	$&$	-	$&$	-	$&$	-	$&$	1_{-0.15}^{+0.06}	$&$	2.33_{-0.42}^{+0.90} 	$&	149.7/134	\\ [0.1cm]
Mrk~509	&$	-10.37 	$&	0.0083	&$	1.73 \pm 0.02	$&$	-	$&$	-	$&$	-	$&$	1.32_{-0.3}^{+0.14}	$&$	3.67_{-1.39}^{+5.56}	$&	313.7/315	\\ [0.1cm]
NGC~2617	&$	-10.63 	$&	0.0026	&$	1.77 \pm 0.014	$&$	-	$&$	-	$&$	-	$&$	-	$&$	-	$&	296.5/318	\\ [0.1cm]
NGC~4151	&$	-9.64	$&	0.0017	&$	1.37_{-0.05}^{+0.09}	$&$	22.1 \pm 0.3	$&$	1.91_{-0.08}^{+0.01}	$&$	0.974 \pm 0.002	$&$	1.77_{-0.05}^{+0.18}	$&$	2.55_{-1.09}^{+1.01}	$&	259.3/267	\\ [0.1cm]
NGC~4593	&$	-10.5	$&	0.0059	&$	1.74 \pm 0.02	$&$	0.72_{-0.15}^{+0.19}	$&$	2.24 \pm 0.14	$&$	1^{\rm fixed}	$&$	-	$&$	-	$&	254/275	\\ [0.1cm]
NGC~5548	&$		$&	0.0034	&$	1.5	$&$		$&$		$&$		$&$		$&$		$&		\\ [0.1cm]
NGC~7469	&$	-10.66	$&	0.0116	&$	1.8 \pm 0.02	$&$	-	$&$	-	$&$	-	$&$	-	$&$	-	$&	190.9/209	\\ \hline
	\end{tabular}

\begin{tablenotes}

\item $^a$ Fluxes are in units of $\rm erg~s^{-1}~cm^{-2}$; $^b$ Column densities are in units of $10^{22}~\rm cm^{-2}$; $^c$ Ionization parameters are in units of $\rm erg~cm~s^{-1}$.
\end{tablenotes}
	
\end{threeparttable}
\end{table*}


We fitted the $0.5-8$~keV spectra using XSPECv12.11.1 \citep{Arnaud96}. We considered a generic model that consists of a power-law component modified by Galactic absorption \citep[{\tt TBabs};][]{wilms00} and ionized absorption intrinsic to the source \citep[{\tt zxipcf};][]{Reeves08}, plus an ionized reflection component using the {\tt Xillver-a-Ec5} table \citep{Garcia2013,Garcia16}. In the XSPEC parlance the model can be written as follows:

$${\tt Model =  TBabs \times (zxipcf \times powerlaw + Xillver)}. $$

\noindent In the case of {\tt Xillver} we fixed the Fe abundance to unity and the inclination angle to 30\degr\ as they could not be  constrained. A simple {\tt TBabs $\times$ powerlaw} model fitted well the spectra of NGC~2617 and NGC~7469. The addition of a reflection component was necessary to fit the Mrk~142 and Mrk~509 spectra. The NGC~4593 spectrum could be fitted well with an ionized absorption only. We needed all spectral components to fit the NGC~4151 spectrum. The best-fit results are listed in Table~\ref{tab:spectra_fits}. 

We note that a detailed spectral study of the X--ray spectra is not necessary in our case. Our aim is to use a simple model to fit the X--ray spectrum in order to compute the $2-10$~keV, unabsorbed, luminosity. The best-fit luminosity values are listed in the third column of Table~\ref{tab:spectra_fits} (in Eddington units). 
\vspace{-15pt}

\section{Time-lags fitting}
\label{sec:lagfit}

In K20, we derived an analytic expression for the time-lags between X-rays and any UV/optical band. In most cases though, the UV/optical time-lags are measured  with respect to a UV band (usually around $\sim 2000$\AA). Using Equation (8) in K20,  we derive  the following analytic expression for the dependence of the time lags between two UV/optical bands on the various parameters of the X--ray corona/accretion disc system:

\begin{eqnarray}
\tau (\lambda) - \tau (\lambda_{\rm R}) &=& A(h_{10}) M_7^{0.7} f_1(\dot{m}_{0.05})f_2(L_{\rm X,0.01}) \nonumber \\
& \times &\left(\lambda_{1950}^{B(h_{10})} - \lambda_{\rm R,1950}^{B(h_{10})} \right)~{\rm day}, 
\label{timelagseq}
\end{eqnarray}

\noindent where $\tau(\lambda)$ and $\tau(\lambda_{\rm R})$ are the time-lags between X--rays and the bands at $\lambda$ and $\lambda_{\rm R}$, respectively ($\lambda_{1950}=\lambda/1950$~\AA, and $\lambda_{\rm R, 1950}=\lambda_{\rm R}/1950$\AA). For each source we adopted the reference band in the original paper where the time-lags were presented, except for NGC~4593 (see Section~\ref{sec:bestfitres}). In the equation above, $h_{10}$ is the height of the lamp-post in units of 10~\rg, $M_7$ is the BH mass in units of $10^7 M_\odot$, $\dot{m}_{0.05}$ is the accretion rate in units of 5$\%$ of the Eddington limit, $L_{\rm X,0.01}$ is the observed, $2-10$~keV luminosity in units of 0.01 of the Eddington luminosity, and, 

\begin{eqnarray}
A(h_{10}) &=& 0.27 + 0.031 h_{10} - 0.0007  h_{10}^2 \\
B(h_{10}) &=& 1.395 - 0.043  h_{10} + 0.0017   h_{10}^2 \label{eq:height0} \\ \label{eq:mdot_a0}
f_1(\dot{m}_{0.05}) &=& 0.636 + 0.4 \dot{m}_{0.05} - 0.059 \dot{m}_{0.05}^2 \\ 
 &+& 0.0033 \dot{m}_{0.05}^3 \nonumber \\
f_2(L_{\rm X,0.01}) &=& L_{\rm X,0.01}^{-0.763} \left[ \frac{1}{2} \left( 1 + L_{\rm X,0.01} ^{0.12}\right) \right]^{14.03}
\end{eqnarray}
\noindent in the case when $a^{\ast}=0$, and,

\begin{eqnarray}
A(h_{10}) &=& 0.164 + 0.039  h_{10} - 0.0012  h_{10}^2 \\
B(h_{10}) &=& 1.346 - 0.037 h_{10} + 0.0013   h_{10}^2 \label{eq:height1} \\ \label{eq:mdot_a1}
f_1(\dot{m}_{0.05}) &=& 0.823 + 0.193 \dot{m}_{0.05} - 0.023 \dot{m}_{0.05}^2 \\
 &+& 0.0012 \dot{m}_{0.05}^3 \nonumber \\
f_2(L_{\rm X,0.01}) &=& L_{\rm X,0.01}^{0.025} \left[ \frac{1}{2} \left( 1 + L_{\rm X,0.01} ^{0.79}\right) \right]^{0.38}
\end{eqnarray}
\noindent in the case when $a^{\ast}=1$.

\begin{figure*}\centering
	\includegraphics[width=0.95\linewidth]{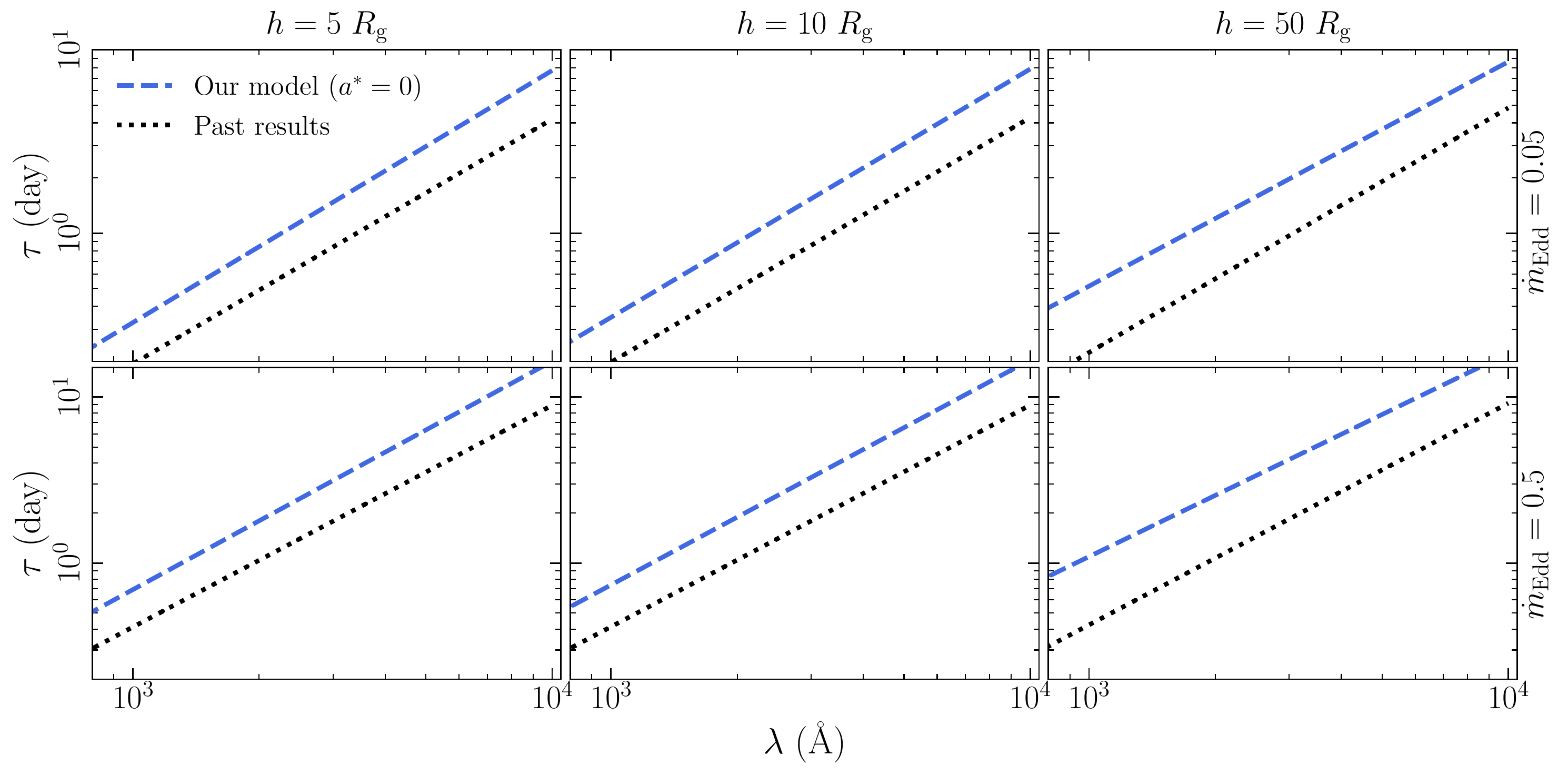}
    \caption{Comparison between the K20 and past time-lag spectra in the case of a non-rotating BH (blue dashed and black dotted lines, respectively). We assume a BH mass of $5\times10^7~M_\odot$, corona heights of 5, 10, and 50~\rg\ (left to right) and accretion rates of \mdot$=0.05$ and 0.5 (top and bottom, respectively; see Section~\ref{sec:comparison} for details).}
    
    \label{fig:model}
\end{figure*}

\begin{figure}\centering
	\includegraphics[width=0.95\linewidth]{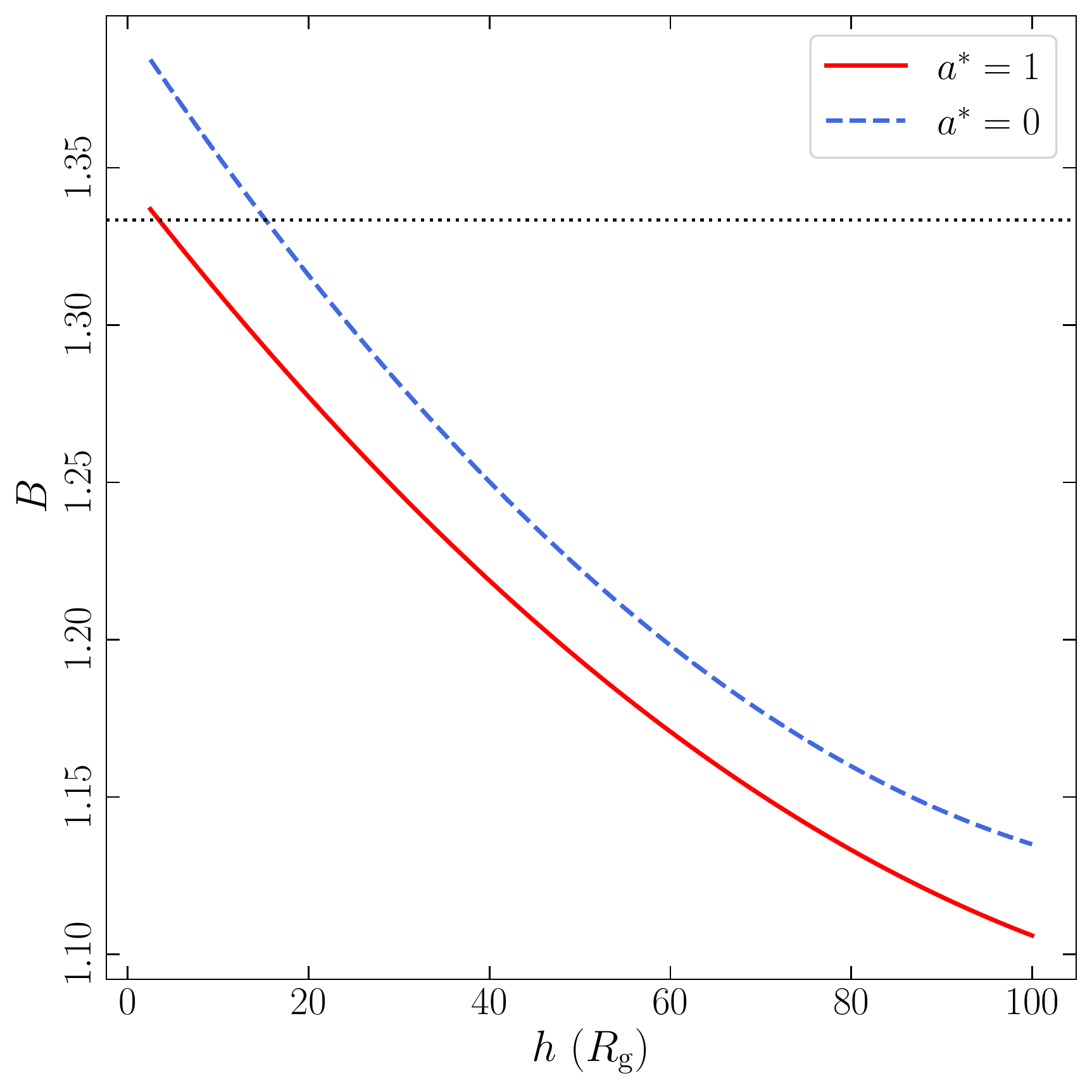}
    \caption{The slope of the time-lag spectra for $a^\ast =0$ and 1 (blue dashed and red solid lines, respectively) given by equations \ref{eq:height0} and \ref{eq:height1}, as function of height. The black dotted line indicates the $4/3$ value that is assumed according to previous models.}
    \label{fig:slope}
\end{figure}
\subsection{Comparison with past results}
\label{sec:comparison}

Like K20, most previous studies assumed a standard accretion disc irradiated by an X--ray source in the lamp post geometry. In most cases, time-lags were supposed to be equal to the light travel time between the X--ray source and a disc radius \citep[see e.g.,][]{Cackett07, Fausnaugh16, Edelson19, Santisteban20}, and they are given by,

\begin{eqnarray}
\tau (\lambda)\!-\!\tau (\lambda_{\rm R})\hspace*{-3mm} &=& \hspace*{-3mm}\frac{1}{c}\!\left( X\frac{k\lambda_{\rm R}}{hc}\right)^{\!4/3}\!\left[ \frac{3G M \dot{M}}{8\pi \sigma}\!+\! \frac{(1-A) L_{\rm X,t}H}{4\pi \sigma}\right]^{\!1/3} \nonumber \\
& &\hspace*{-3mm}\times\left[ \left(\frac{\lambda}{\lambda_{\rm R}}\right)^{\!4/3} - 1\right]~{\rm sec},
\label{eq:timelagOLD}
\end{eqnarray}

\noindent where $\lambda_{\rm R}$ is the reference band, $A$ is the disc albedo, $L_{\rm X,t}$ is the total X-ray luminosity, $M$, $\dot{M}$ and $H$ are the BH mass, accretion rate, and X--ray source height in S.I. units. The  constant $X$ takes the value of 4.97 if the total flux at $\lambda$ is emitted from a single annulus of radius defined according to Wien's law. However, $X$ is smaller ($X \simeq 2.49$) if the characteristic disc radius for the flux at $\lambda$ is a flux weighted mean of all radii that contribute to it. We note that $L_{\rm X,t}$ is difficult to determine in practice. If $\Gamma\geq 2$, then it is crucial to know the low-energy rollover, which is determined by the typical energy of the seed photons, as seen by the corona, and is difficult to detect. At flatter slopes, it is crucial to know the high-energy cutoff (during the time of the monitoring observations). For that reason, K20 used the (easily determined) $2-10$~keV luminosity instead (see Equation~\ref{timelagseq}). 

To compare this model to ours, we consider the $a^\ast = 0$ case for $M_{\rm BH} = 5\times 10^7~M_\odot$, \mdot$=0.05$ and 0.5, and $h=5, 10$ and 50~\rg. We also assumed $A=0.2$, $\Gamma =2$, a high-energy cutoff of 300~keV, and a $2-10$~keV luminosity which is consistent with the assumed accretion rate \citep[according to][ see their Figure~10 and Table~3]{Lusso12}. Then we computed the corresponding total X-ray luminosity by estimating the appropriate low-energy rollover (see Equation~5 and Figure~1 in K20). We note that to estimate $\dot{M}$ in S.I. units (as required in Equation~\ref{eq:timelagOLD}), we adopted a radiative efficiency\footnote{We remind the reader that the accretion rate in physical units is defined as $\dot{M} = L/\eta c^2$, where $\eta$ is the radiative efficiency and it is a function of the BH spin, increasing from 0.057 for $a^\ast =0$ to 0.42 for $a^\ast = 1$.} of $0.057$, as is appropriate for a non-rotating BH.  

Figure~\ref{fig:model} shows the K20 and the previous time-lags models, in the case of $X=4.97$. The latter model predicts a constant slope of 4/3, while our model predicts a height-dependent slope. The difference in slopes is more pronounced at large heights, but it is small and may not be easy to detect in practice. To further illustrate the differences in slope, we plot in Figure~\ref{fig:slope} the K20 slopes for $a^\ast =0$ and 1, as a function of height. The horizontal, black dotted line in the same figure indicates the 4/3 slope. The figure shows that difference between the K20 model slopes and 4/3 increases with increasing $h$, reaching a maximum of $\sim 20~\%$, at $h=100$~\rg. However, the main discrepancy between the K20 and the time-lags defined by Equation~\ref{eq:timelagOLD} is in their amplitude. Our model predicts lags that are $\sim 2$ times larger than the model in Equation~\ref{eq:timelagOLD}, for $X=4.97$ (the amplitude difference would be larger for $X=2.49$). 

In K20, we did not compute time-lags as the light travel time between the X--ray source and a particular disc radius. Instead, we computed the disc response to X--rays by considering in detail the incident X--ray spectrum on each disc element (which depends on general relativity effects), and the disc ionisation in computing the disc reflection spectrum. The resulting model time-lags are based on the width of the transfer functions at each wavelength (see Equation 6 in K20). We believe that the main difference between the K20 and previous models is primarily due to the fact that the K20 disc transfer functions are wider than what was thought in the past. 

A second major difference between K20 and the modelling of the time-lags in the past is the effects of BH spin. K20 also considered the case of a maximally rotating BH ($a^\ast = 1$). They found that the $a^\ast=0$ and $a^\ast=1$ time-lags will be (very) different, mostly in their normalization, even if all the other parameters are the same. This is mainly because the BH spin determines the value of the accretion rate in  physical units ($\rm M_\odot~yr^{-1}$). For a given accretion rate in Eddington units, the accretion rate in physical units is higher for a low spin, and the disc is hotter. Consequently, the disc response functions will be broader, and the time-lags will be larger when $a^\ast=0$ (e.g., see the discussion in Sections 3.2 and 3.4, and figures 16,17, 18 and 23 in K20). It is for this reason that the K20 models have different normalisations for a non-rotating and a maximally rotating BH. It is exactly because of this result that model fitting of the observed time-lags spectra can be used to determine the BH spin, when combined with spectral information as well (as we show below). 

\begin{figure*} \centering
	\includegraphics[width=0.28\linewidth]{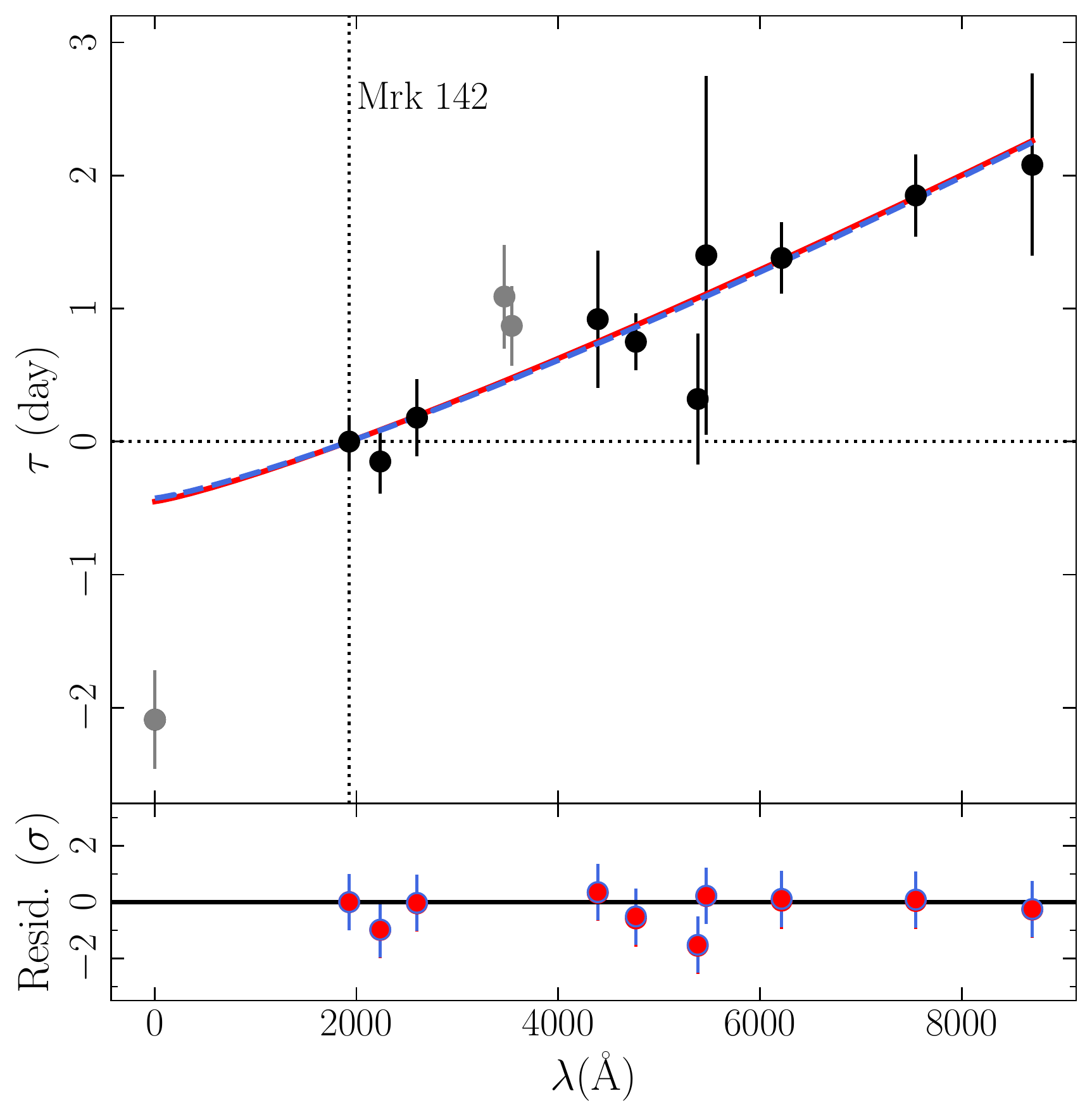}
	\includegraphics[width=0.28\linewidth]{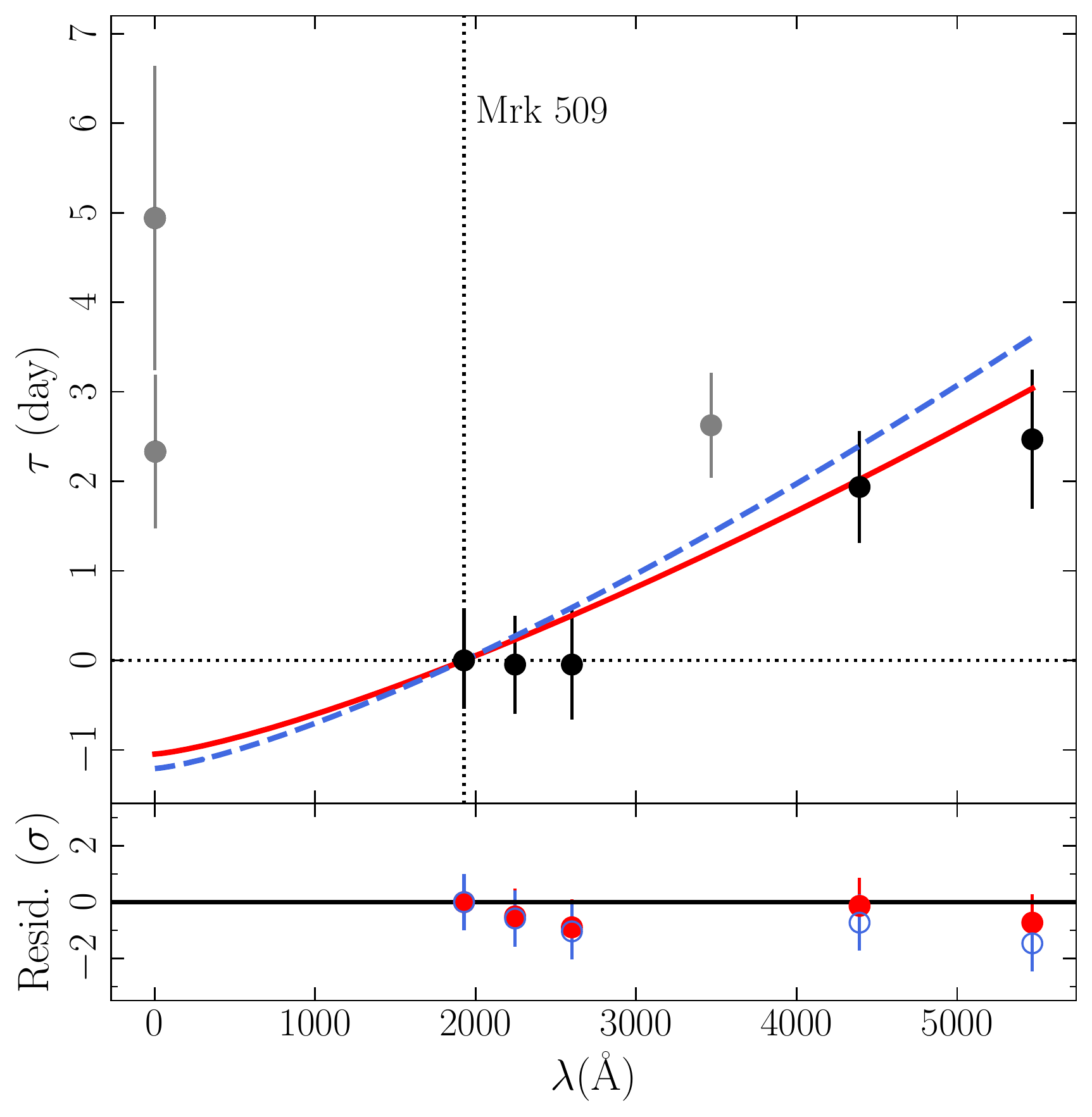}
	\includegraphics[width=0.28\linewidth]{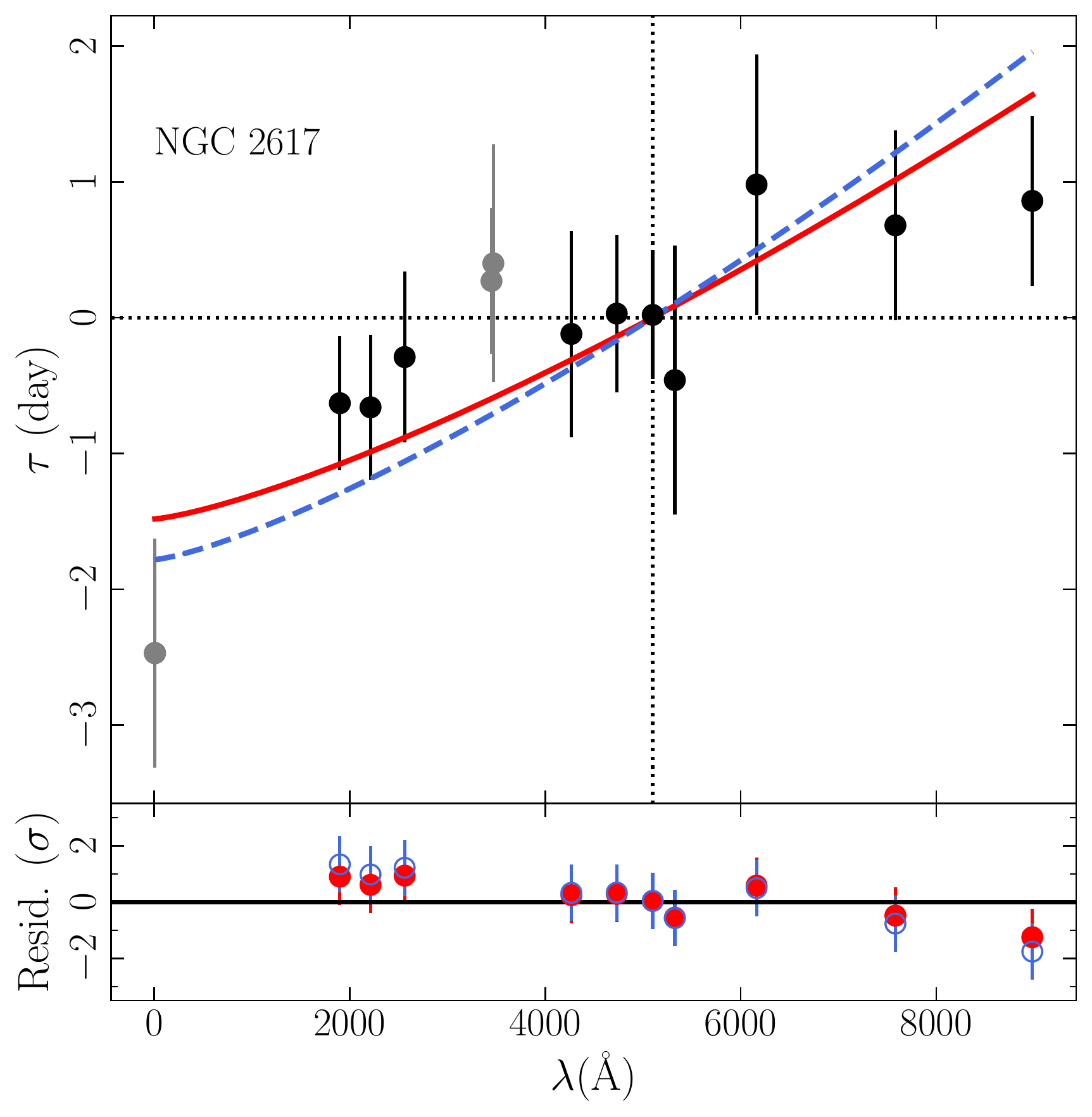}\\
	\includegraphics[width=0.28\linewidth]{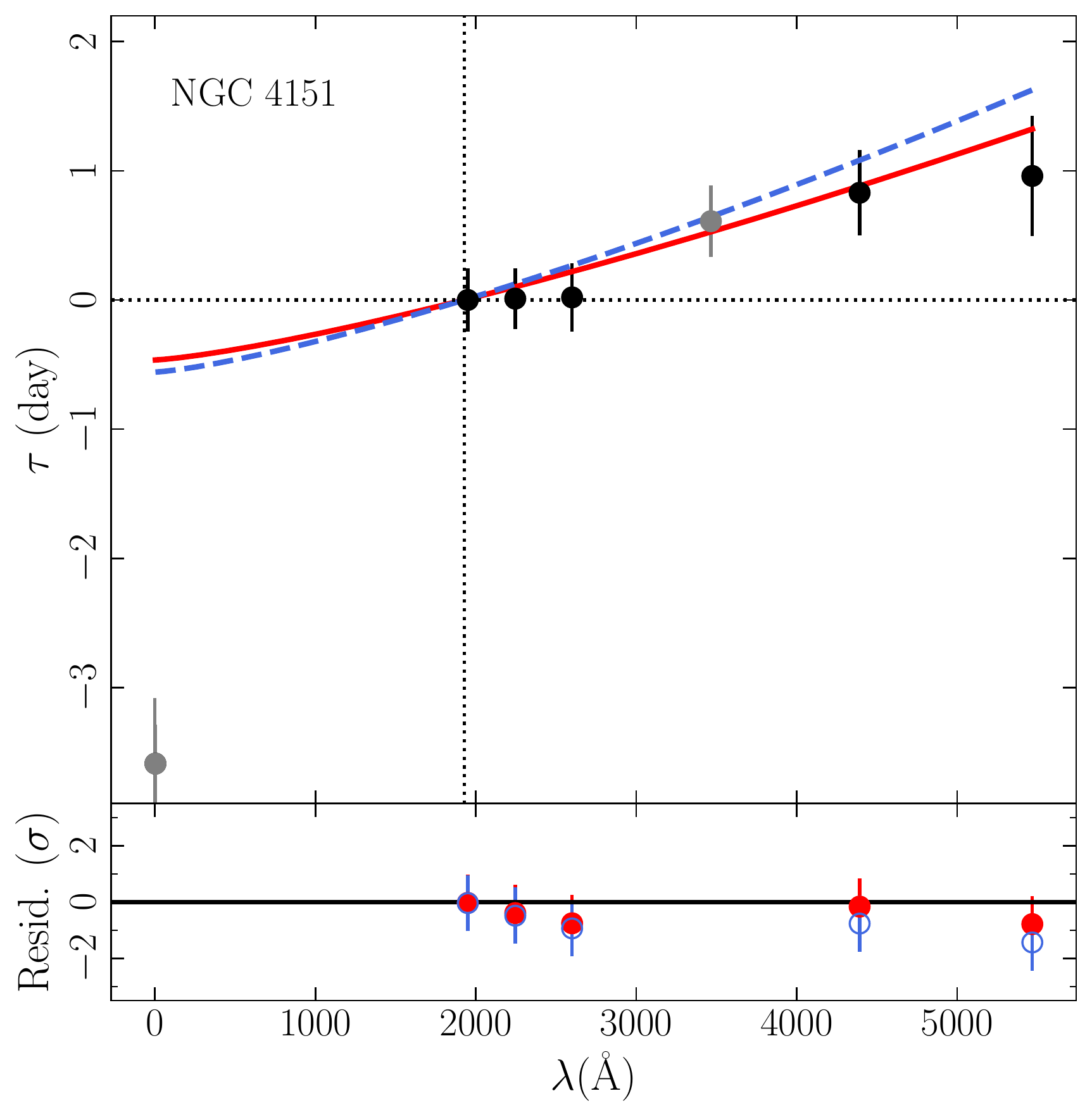}
	\includegraphics[width=0.28\linewidth]{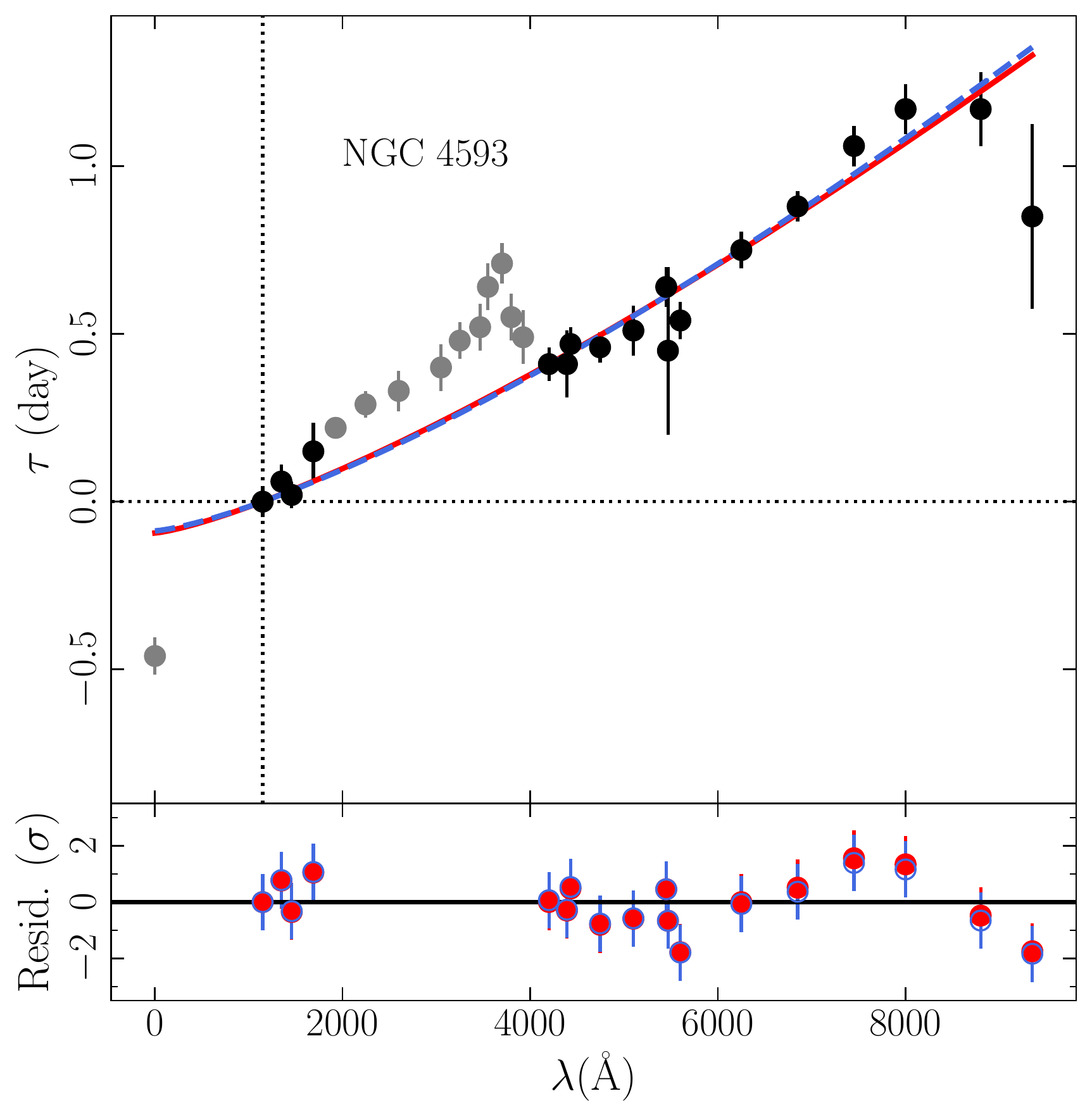}
	\includegraphics[width=0.28\linewidth]{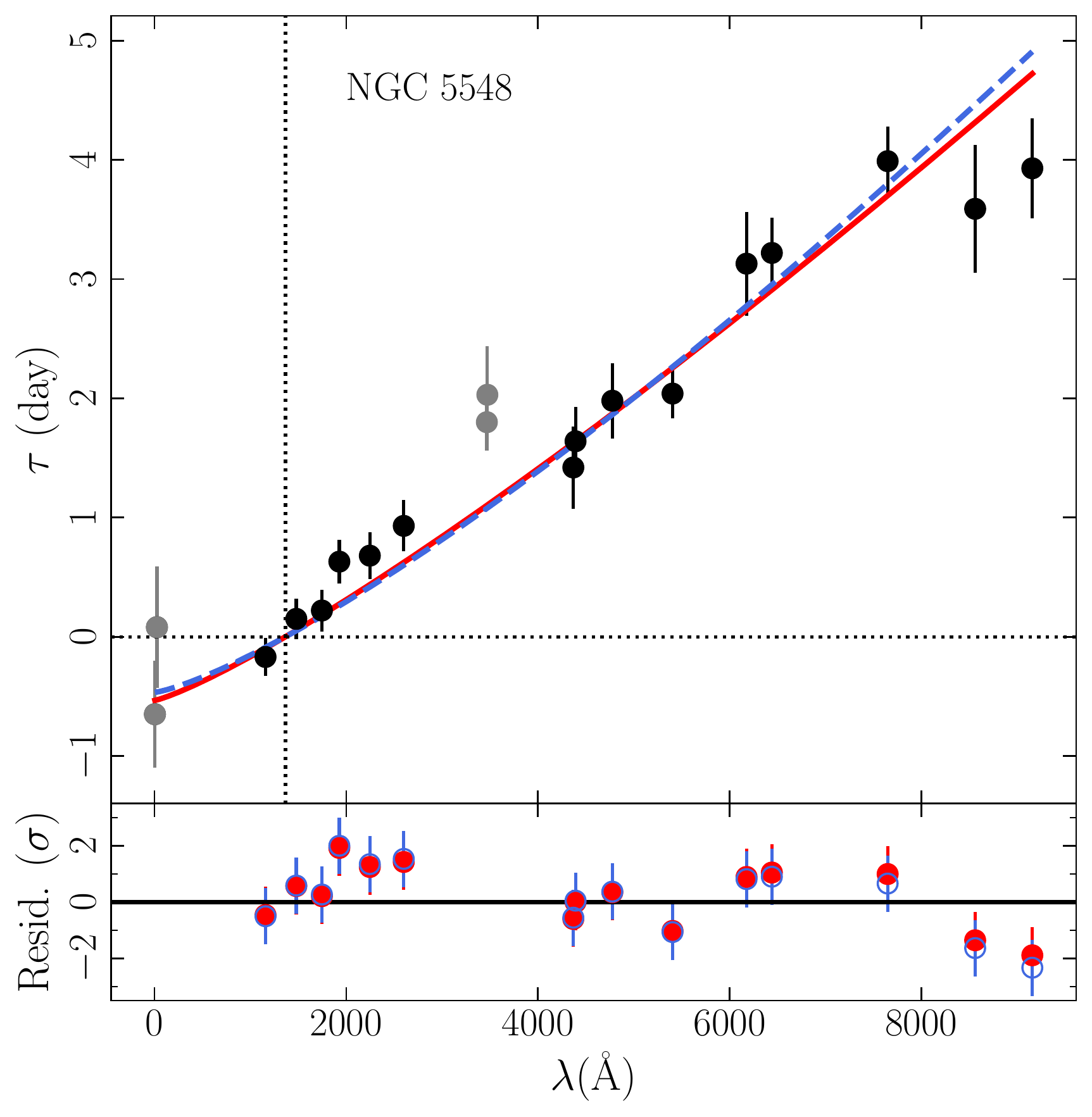}\\
    \includegraphics[width=0.28\linewidth]{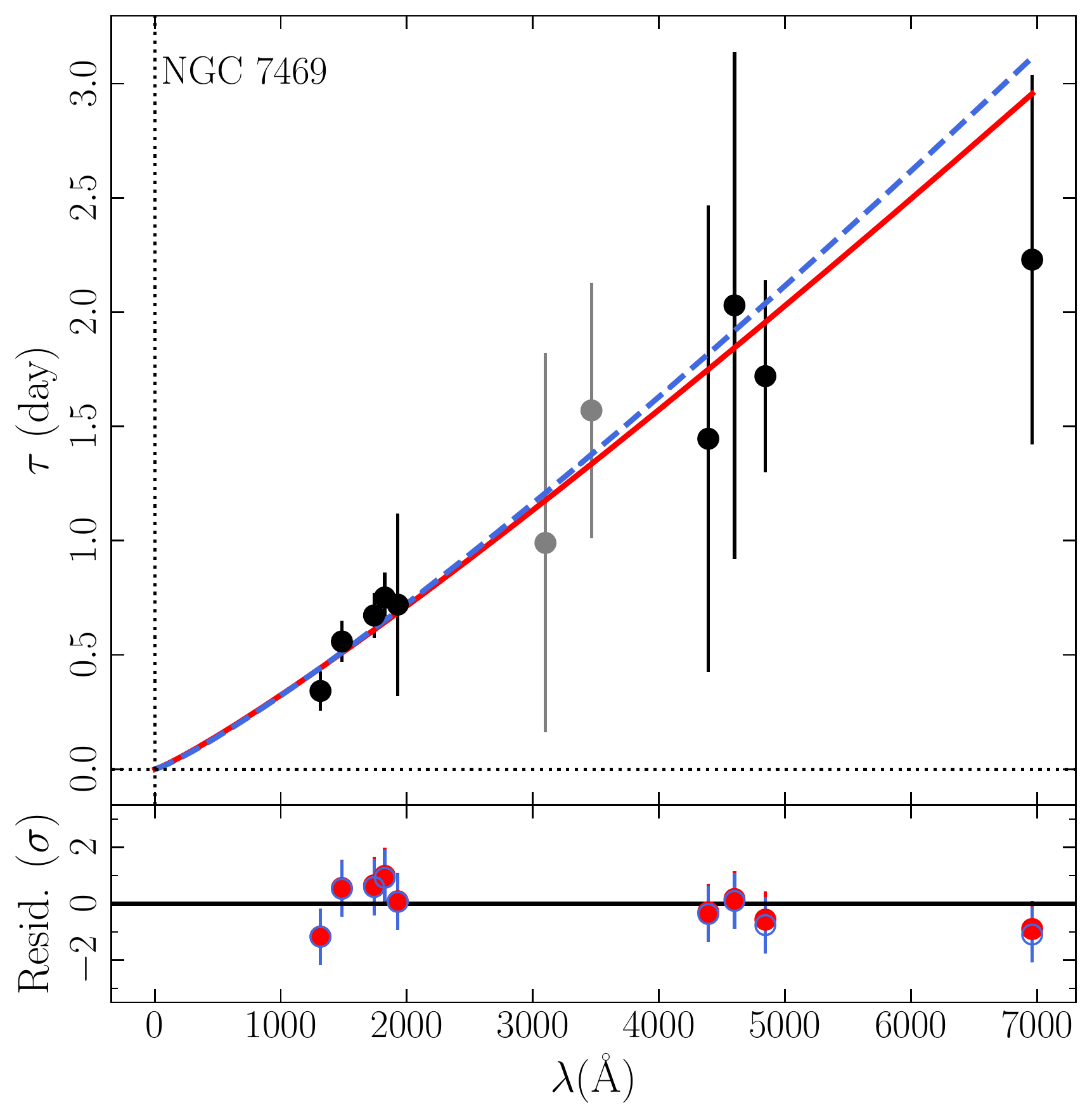}
   
    \caption{The time-lag spectra and the best-fit models for $a^\ast = 0$ and 1 (dashed blue and solid red lines, respectively).  Grey points indicate time-lags excluded from the fit. Bottom panels show the corresponding residuals.}
    \label{fig:lags}
\end{figure*}

\subsection{The best-fit results}
\label{sec:bestfitres}

We used a Markov chain Monte Carlo (MCMC) approach to fit the time-lags spectra using the {\tt emcee}\footnote{\url{https://emcee.readthedocs.io/en/stable/}} sampler \citep{Emcee13}. We fixed $M_{\rm BH}$ and $L_{\rm X,Edd}$ to the values listed in Tables \ref{tab:sources} and \ref{tab:spectra_fits}, and we left the height and the accretion rate as free parameters. We assumed flat priors in the range of $[2.5~ \rm R_g, 100~ R_g]$ and $[0.005, 1.5]$ for  $h$ and \mdot, respectively. We fitted the time-lags twice, for $a^\ast = 0$ and $a^\ast = 1$. Each time we performed the MCMC fit using 100 walkers and chains that are $15,000$ long, discarding the first $5,000$ steps as part of the `burn-in' phase. 

The best-fit results are listed in Table~\ref{tab:lag_fits} (errors correspond to $1\sigma$ confidence intervals). The model fits well the time-lags spectra of all sources. In fact, the reduced $\chi^2$ in some cases is smaller than one, probably due to the large errors of the observed time-lags. We used time-lags estimated following the interpolated cross-correlation function \citep[ICCF;][]{White94}, with errors  computed using the flux randomization (FR) and random subset sampling (RSS) approach \citep{Peterson2004}. They are probably overestimated, as the NGC~4593 case indicates. The $\chi^2$ statistics are more reasonable for this source, where we use the FR only lag uncertainties of \cite{Cackett18}. JAVELIN \citep{Zu11, Zu13} usually provides smaller time-lags uncertainties \citep[see e.g., discussion in][]{Cackett20, Yu20}. However, at least one of the assumptions of this method (i.e., the top-hat transfer function) is not compatible with the results of K20, so we decided to use the time lags estimates based on ICCF.

Figure~\ref{fig:lags} shows the time-lags and the best-fit models. Vertical dotted lines indicate the reference band. Grey points show the time-lags we excluded from the fit. They include the X--ray time lags and the time-lags between the reference and the $U-$bands. The latter usually lie above the best-fit models, probably due to significant continuum emission from the Broad Line Region \citep[BLR;][]{Korista01, Korista19, Lawther18}. This effect is particularly evident in NGC~4593 where the time-lags down to $\sim 2000$~\AA\ may be affected by the BLR continuum emission \citep{Cackett18}. We chose $\lambda_{\rm R}=1150$~\AA\ in this object because the original reference band, {\it UVW2}, may be affected by excess emission from the BLR. The best-fit residual plots indicate a similar effect in NGC~2617 and NGC~5548. Time lags between $2000-3500$~\AA\ may be affected by excess contribution from the BLR, but we cannot be certain because of the large time-lags errors. For consistency, we exclude all the $U/u-$band data points from our fits for the rest of the sources.

\begin{figure*}
	\includegraphics[width=0.28\linewidth]{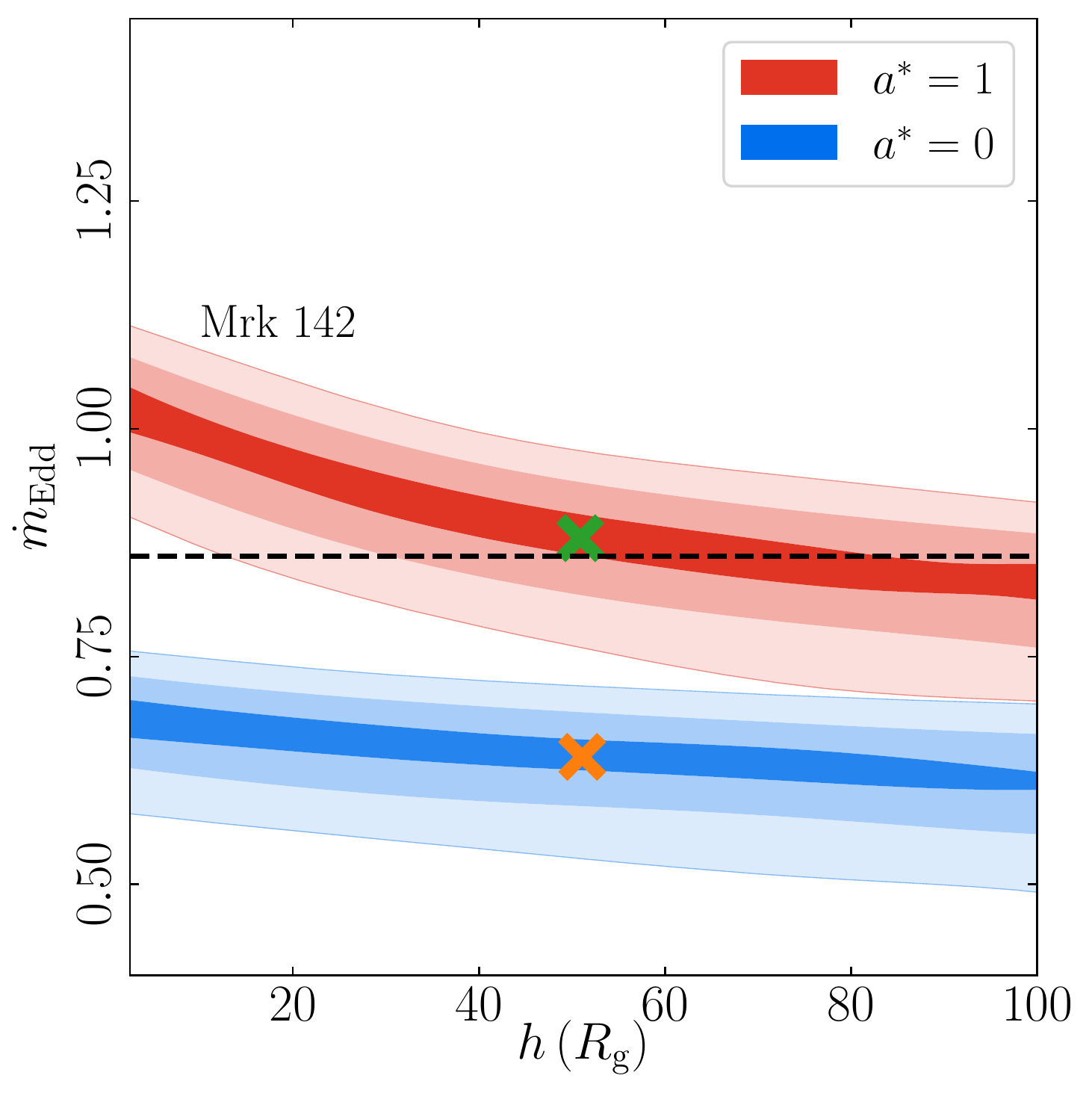}
	\includegraphics[width=0.28\linewidth]{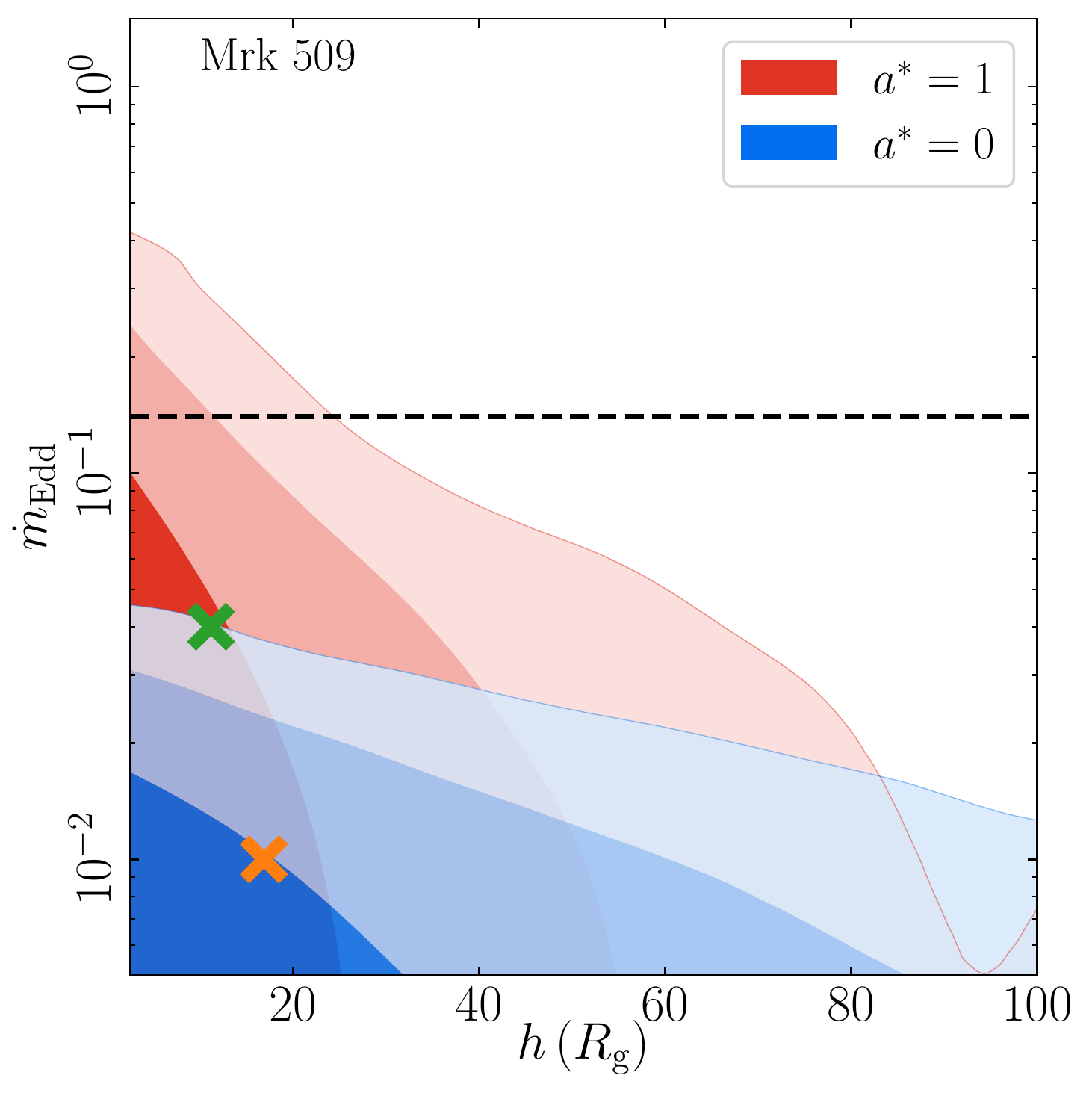}
	\includegraphics[width=0.28\linewidth]{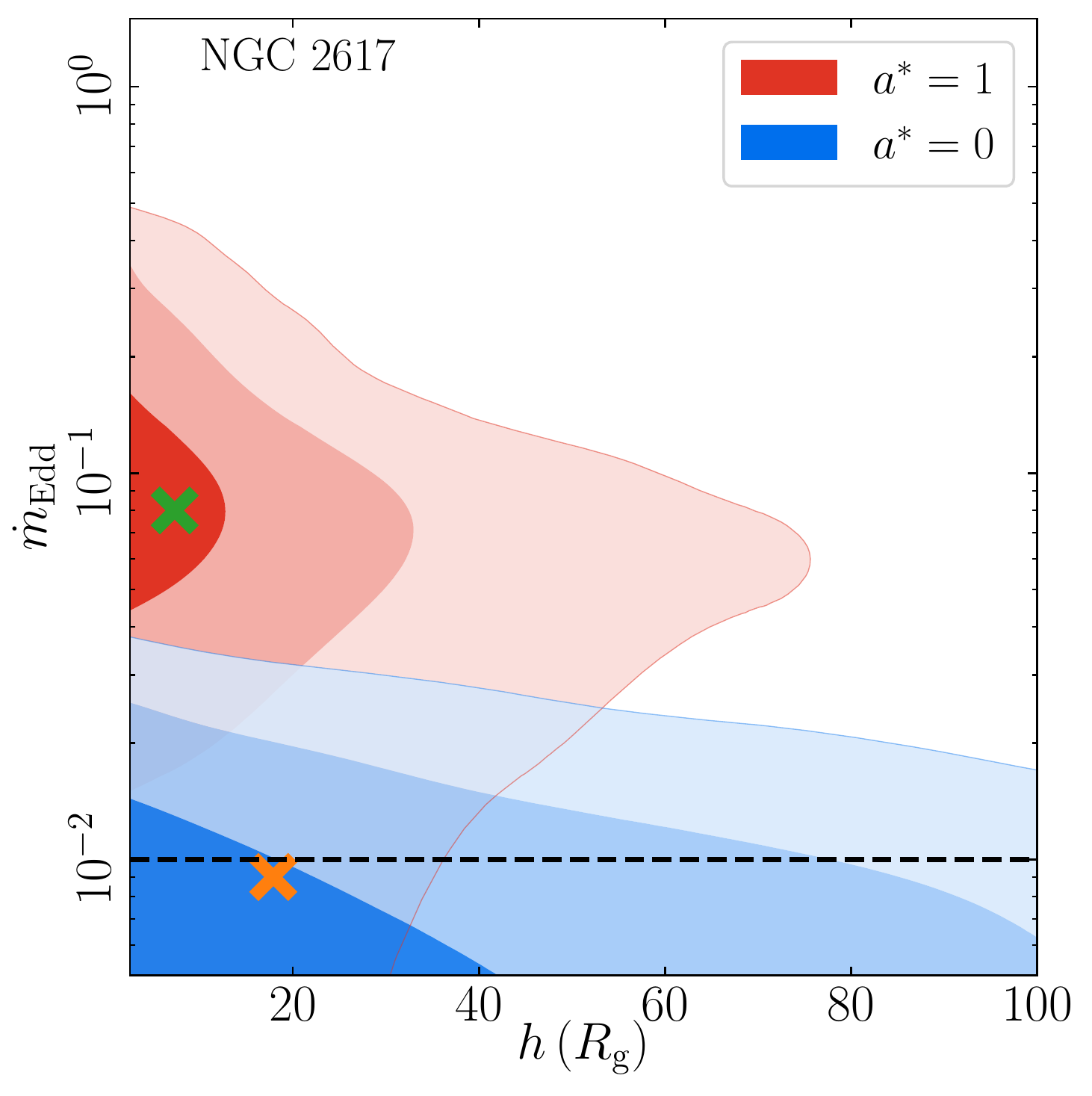}\\
	\includegraphics[width=0.28\linewidth]{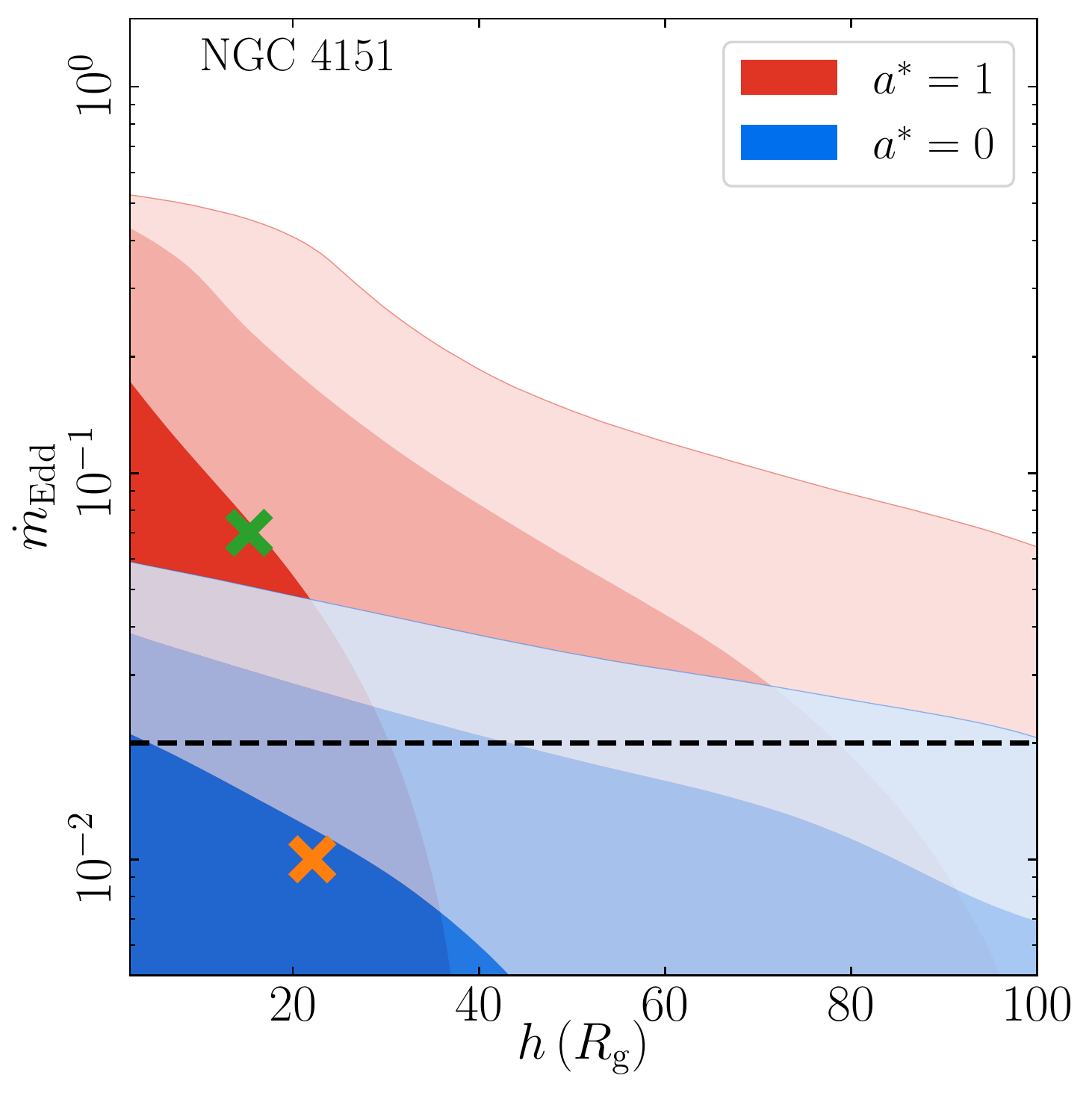}
	\includegraphics[width=0.28\linewidth]{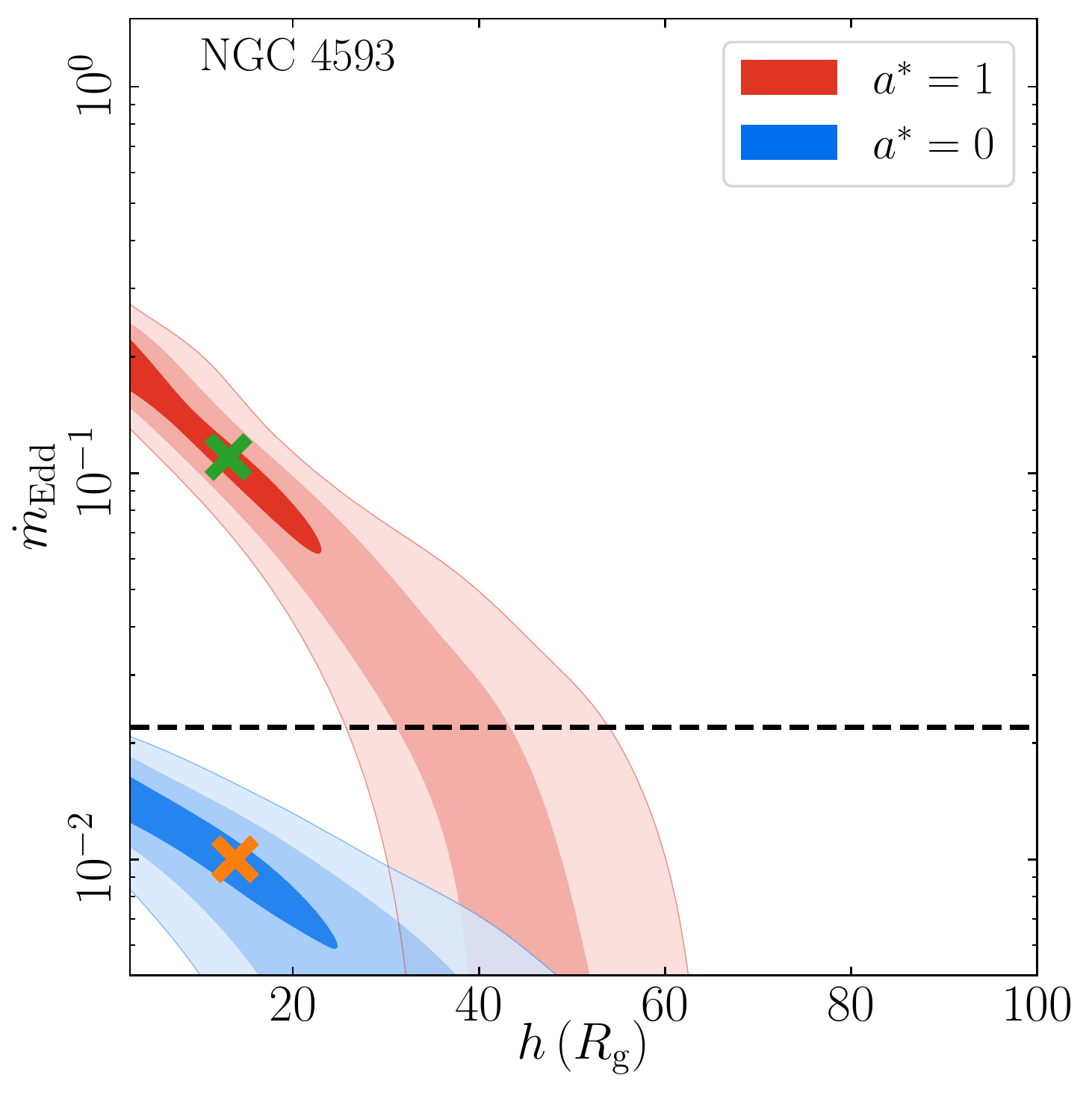}
	\includegraphics[width=0.28\linewidth]{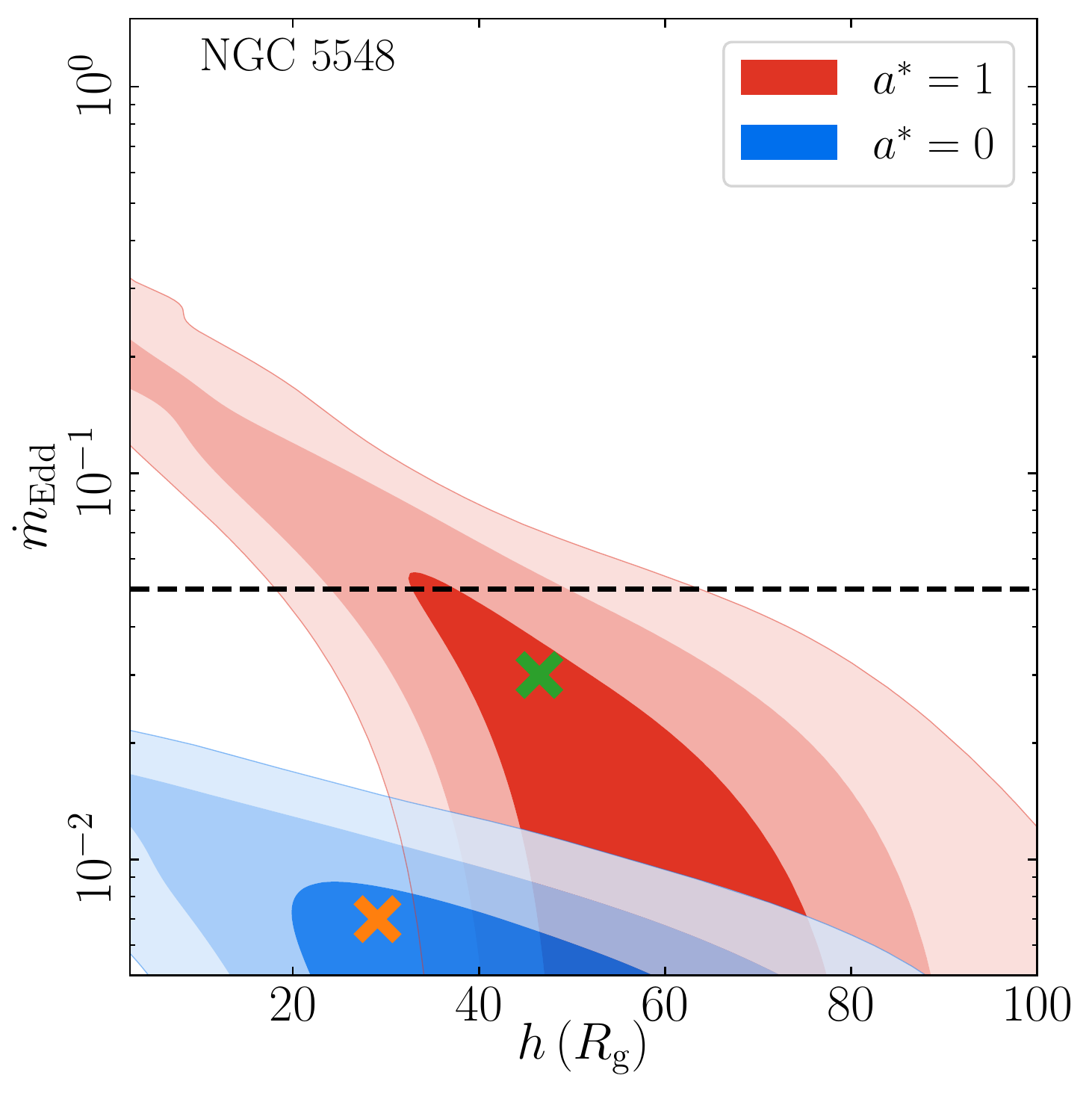}\\
    \includegraphics[width=0.28\linewidth]{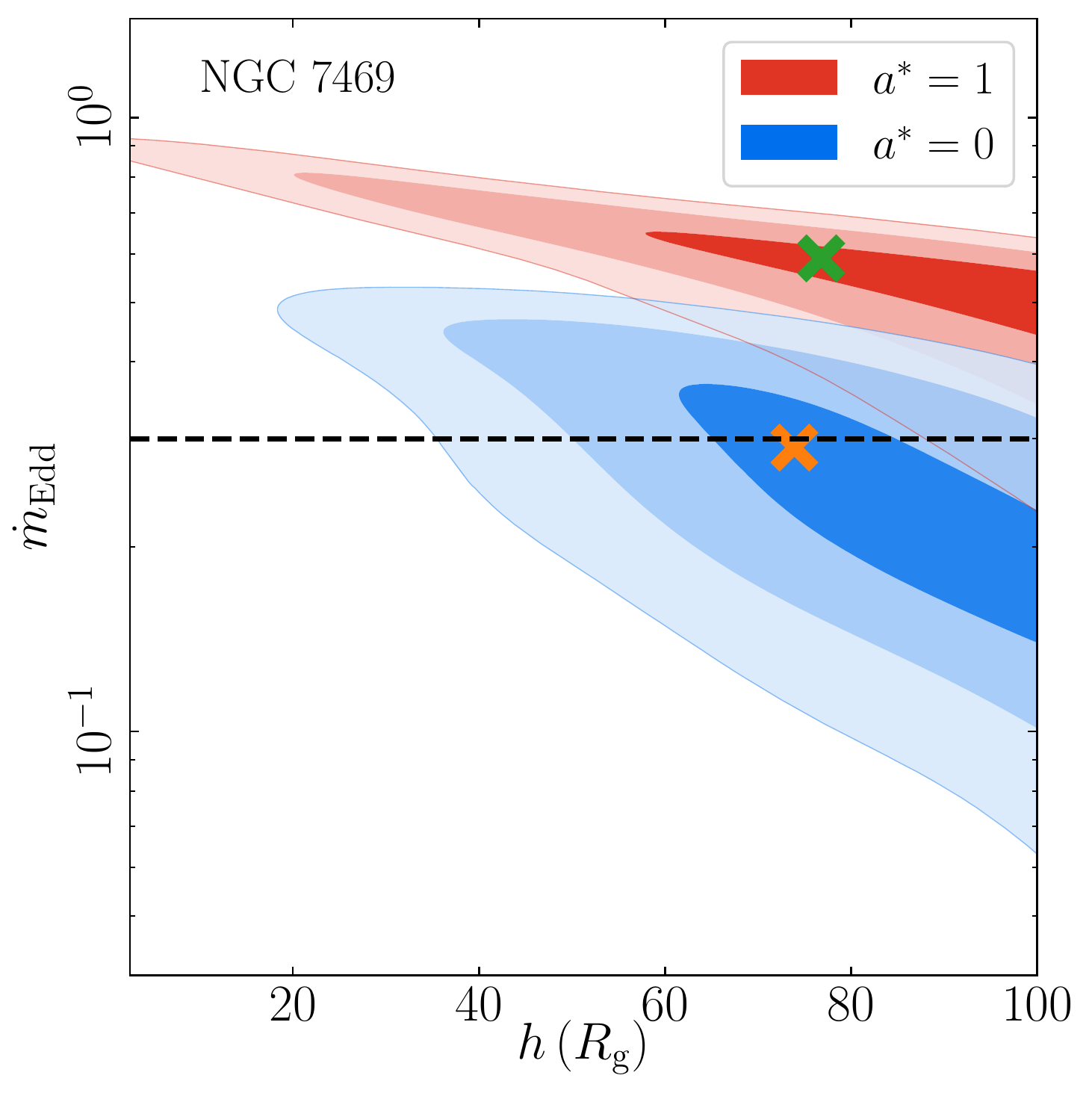}
    
    \caption{\mdot\ vs height confidence contours ($1,2,$ and $3\sigma$) obtained from the MCMC fitting for $a^\ast = 0$ (blue) and 1 (red). The orange and green crosses indicate the best-fit results for $a^\ast =0$ and 1, respectively. The dashed horizontal lines show the \mdot\ values reported in the literature.}
    \label{fig:mcmc}
\end{figure*}

Figure~\ref{fig:mcmc} shows the ``\mdot\ vs height'' confidence contours obtained from the MCMC fitting for $a^\ast = 0$ and 1 (blue and red contours, respectively). Strong to light colours indicate the $1,2,$ and $3\sigma$ confidence contours. The horizontal, dashed lines indicate the accretion rate values from the literature (listed in Table~\ref{tab:sources}). The widest contours are those for NGC~4151 and Mrk~509. Their time-lags spectra have the smallest number of points, and large uncertainties. The opposite is true for NGC~4593 and NGC~5548, to a lesser extent. The time-lags are estimated in many wavebands, and the errors are smaller. Clearly, time lags spectra with many points, and small errors are necessary to constrain theoretical models and determine the height of the X--ray corona and the accretion rate accurately. 

\begin{table}
	\caption{The Best-fit results, for $a^\ast = 0$ and 1. We list the $1\sigma$ errors for the best-fit parameters.} 
	\label{tab:lag_fits}
\resizebox{\columnwidth}{!}{
\begin{tabular}{lllll} 
\hline
Source	&$		$&$	h~(\rm R_g)	$&$	\dot{m}_{\rm Edd}	$&	$\chi^2$/dof	\\ \hline
Mrk 142			&$	a^\ast = 0	$&$	51.1 ~[2.5, 100]	$&$	0.64~ [0.59, 0.70]	$&	3.7/8	\\ [0.1 cm]
			&$	a^\ast = 1	$&$	50.9 ~[2.5, 100]	$&$	0.88~[0.81, 1.03]	$&	3.9/8	\\ [0.1 cm]
Mrk 509			&$	a^\ast = 0	$&$	16.9 ~[2.5, 31.6]	$&$	0.01 ~[0.005, 0.016]	$&	4.1/3	\\ [0.1 cm]
			&$	a^\ast = 1	$&$	11.2 ~[2.5, 25.3]	$&$	0.04~[0.005, 0.10]	$&	1.6/3	\\ [0.1 cm]
NGC 2617			&$	a^\ast = 0	$&$	17.9 ~[2.5, 42.3]	$&$	0.009 ~[0.005, 0.014]	$&	8.7/8	\\ [0.1 cm]
			&$	a^\ast = 1	$&$	7.3 ~[2.5, 12.9]	$&$	0.08~[0.04, 0.16]	$&	4.7/8	\\ [0.1 cm]
NGC 4151			&$	a^\ast = 0	$&$	22.1 ~[2.5, 43.2]	$&$	0.01 ~[0.005, 0.02]	$&	3.7/3	\\ [0.1 cm]
			&$	a^\ast = 1	$&$	15.3 [2.5, 37.1]	$&$	0.07 ~[0.005, 0.17]	$&	1.4/3	\\ [0.1 cm]
NGC 4593			&$	a^\ast = 0	$&$	13.8~[2.5, 24.8]	$&$	0.01~[0.006, 0.016]	$&	14.2/16	\\ [0.1 cm]
			&$	a^\ast = 1	$&$	13.1~[2.5, 23.3]	$&$	0.11~[0.06, 0.22]	$&	14.7/16	\\ [0.1 cm]
NGC 5548			&$	a^\ast = 0	$&$	29.1 [23.2, 58.5]	$&$	0.007 ~[0.005, 0.008]	$&	20.2/12	\\ [0.1 cm]
			&$	a^\ast = 1	$&$	46.5 ~[32.7, 77.7]	$&$	0.03 ~[0.005, 0.05]	$&	17.5/12	\\ [0.1 cm]
NGC 7469			&$	a^\ast = 0	$&$	73.9 ~[61.5, 100]	$&$	0.29~[0.13, 0.35]	$&	4.8/7	\\ [0.1 cm]
			&$	a^\ast = 1	$&$	76.8 ~[58.2, 100]	$&$	0.59 ~[0.44, 0.65]	$&	4.3/7	\\ [0.1 cm]
		\hline
	\end{tabular}
	}
\end{table}


\begin{figure}\centering
	\includegraphics[width=0.95\linewidth]{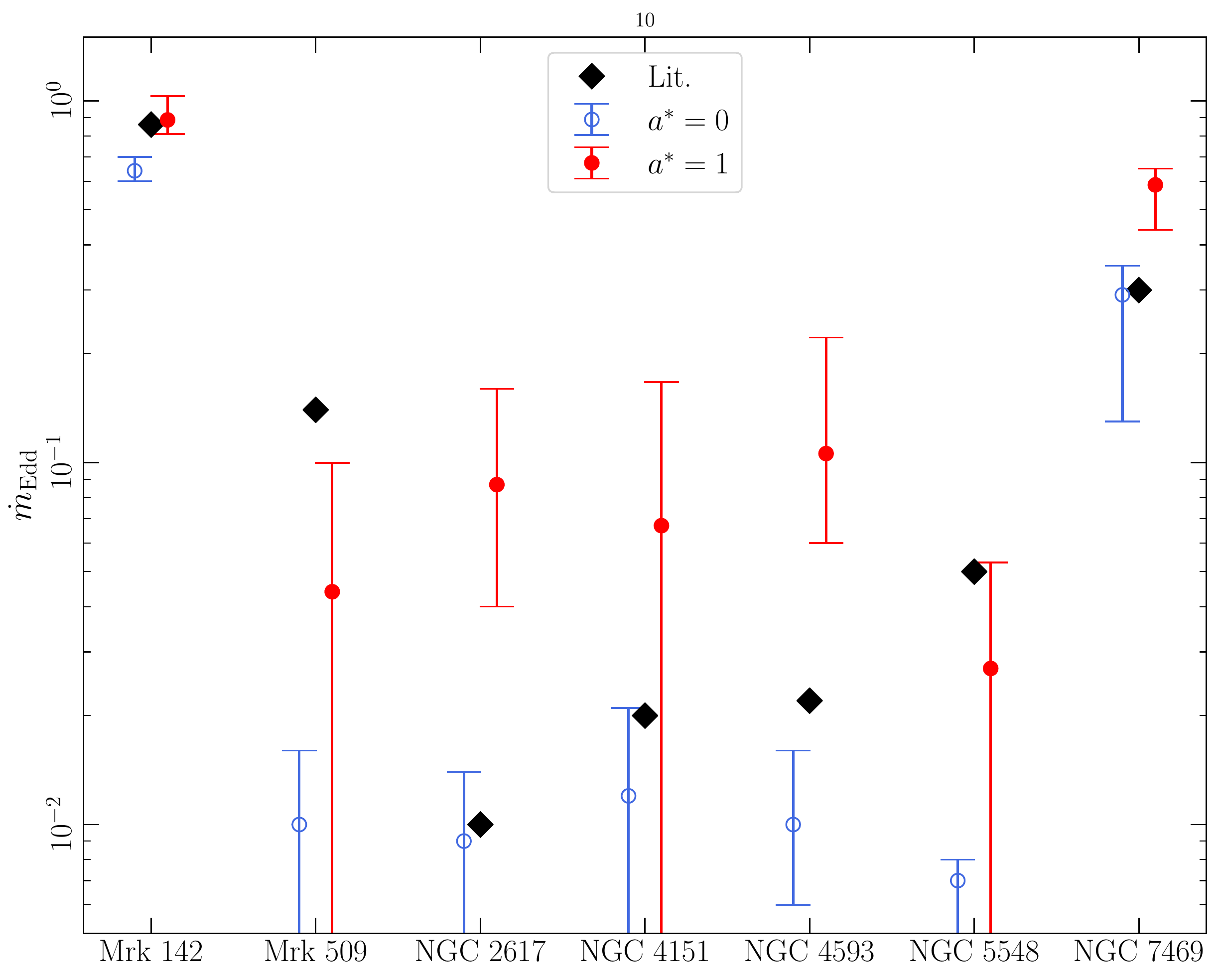}
    \caption{The measured \mdot\ for $a^\ast = 0$ and 1 (open and filled circles, respectively). Black diamonds show the \mdot\ values reported in the literature.}
    \label{fig:mdot}
\end{figure}

Nevertheless, we can still use the best-fit results to constrain the accretion flow and the X-ray geometry in these AGN. Open and filled circles in Figure~\ref{fig:mdot} show the best-fit accretion rates from our fits to the time lags spectra in the case of a Schwarzschild  and a maximally rotating BH ($a^\ast=0$ and 1, respectively). These are estimates of the power that heats the accretion disc. Filled diamonds in the same figure indicate the accretion rates listed in Table~\ref{tab:sources}. They are based on $L_{\rm Bol}/L_{\rm Edd}$ ratio estimates,  and are representative of the total power released by the accretion process. 

A fraction of the power released during the accretion process, say $f$, must be transferred and dissipated in the X-ray corona. The fraction $f$ should not be very large. For example, the 2--10 keV X--ray luminosity is usually less than 10 per cent of $L_{\rm Bol}$ \citep[e.g.,][]{Lusso12}. This should also hold for the total X--ray luminosity when $\Gamma \geq 2$, irrespective of the high energy cut-off. Furthermore, observational evidence indicates that the total X--ray luminosity is approximately half $L_{\rm Bol}$ \citep{Lubinski16}. 

Within our model, where the X--ray source is compact and located above the BH, the power that heats the corona could be the power that is released by the accretion process within a certain radius (say $R_{\rm in})$. Using the {\tt KYNBB}\footnote{\url{https://projects.asu.cas.cz/stronggravity/kyn\#kynbb}} model, we estimate that half of the total disc luminosity is released within $R_{1/2} = 36$ and 11~\rg\ for $a^\ast=0$ and 1, respectively,  assuming an inclination of 30\degr.\footnote{While for smaller inclinations this radius increases by less then $3\,$\rg, for higher inclinations it becomes much smaller. One could also use the estimate of the radius where the matter in Keplerian disc loses half of the total transformed energy when it falls down to the black hole. In that case the radius will be even smaller, $R_{1/2} = 16$ and 2.1~\rg\ for $a^\ast=0$ and 1, respectively.} Thus, if the total power released within this $R_{\rm 1/2}$ could be transferred to the corona then $f=0.5$, and the inner disc will be colder, and would  contribute less to UV/optical. K20 have studied the case where the disc does not emit below a given $R_{\rm in}$, and they have shown that the model time-lags are almost not affected at all, as long as $R_{\rm in} \leq 100$~\rg. This implies that, even if the inner disc does not contribute to the observed UV/optical emission, the time-lags fitting can provide the intrinsic accretion rate.

Based on the discussion above, the data in Figure~\ref{fig:mdot} suggest that NGC~2617 and NGC~7469 host a slowly rotating BH, because the \mdot\ we get when $a^\ast = 1$ is larger than the accretion rate from literature (this is a $2\sigma$ effect) . On the other hand the BHs in Mrk~142, Mrk~509, NGC~5548 and NGC~4593 are most likely rotating fast. The accretion rate we compute when $a^\ast = 0$ is significantly smaller than the accretion rate estimates from the literature (the difference is significant at the $3\sigma$ level in all sources as shown in Figure~\ref{fig:mcmc}). On the other hand, the \mdot\  estimate for $a^\ast=1$ is consistent (within the $1 - 2\sigma$ confidence level)  with the values listed in Tab~\ref{tab:sources}. Finally, both the $a^\ast=0$ and $a^\ast=1$ provide \mdot\ estimates that are consistent with the value listed in Table~\ref{tab:sources} for NGC~4151.

\section{Discussion and Conclusions}
\label{sec:discussion}

We used the K20 model and we fitted the continuum, UV/optical time lags in seven Seyfert galaxies. Time-lags in these sources have been computed using data from recent, intensive,  multi-wavelength observational campaigns. Arguably, these observations have provided the best data sets we have got at the moment in order to perform disc reverberation mapping. K20 took into account all the relativity and disc ionization effects, and they provided an analytic expression for the time-lags as a function of $M_{\rm BH}$, \mdot, the observed X--ray luminosity, the X--ray source height and the BH spin. The K20 model time-lags are fully representative of the disc thermal reverberation time-lags in the case of the lamp-post geometry. 

Our main result is that the K20 time-lags model can fit well the UV/optical time-lags in all sources, with accretion rates which are fully compatible with the accretion rates that have been reported for the AGN in our sample. The presumed discrepancy between the observed and the model UV/optical time-lags resulted in the development of additional models which tried to explain the observed time-lags spectra, e.g., \cite{Cai20}, whose model is based on disc turbulence, and \cite{Gardner17}, who suggested that the UV/optical lags of NGC~5548 are not indicative of the light travel time from X-ray or far-UV reprocessing, but instead could be representative of the time-scale for the outer disc vertical structure to respond to the changing far-UV illumination. However, contrary to previous claims, we find that the magnitude of the observed UV/optical time lags, in all sources, is exactly as expected in the case of a standard accretion disc in the lamp-post geometry, given the BH mass and accretion rate estimates for these objects. 

Our results support the hypothesis of the so-called lamp-post geometry. However, in reality, the corona must have a finite size. It is well possible that the point source corona toy model that we use in our work is good enough to describe a large variety of corona geometries as long as the corona is compact enough in comparison with the overall size of the accretion disc. In that case, the height could be indicative of the position and the size of the corona (e.g., for centrally located spherical corona). Additional work would also be needed to investigate whether extended coron\ae\ \citep[e.g.,][]{Wilkins2016} are consistent with the observations. We do not believe that our results (which are consistent with a compact corona) necessarily exclude the possibility of an extended corona. It is necessary to model the time-lags when the corona is extended, and fit them to the data.

In the following section we compare our results with past $\dot{m}$ measurements, and we discuss the implications regarding BH spin. We would like to comment at this point on the accretion rate estimates for Mrk~142 and NGC~7469. We measure a high accretion rate for both objects, in agreement with past measurements. We find it very interesting that completely different methods suggest a similarly high accretion rate for these two objects. We believe this agreement supports the validity of our results. It has been suggested that the accretion disc structure changes at high accretion rates. Radiation pressure may dominate the accretion flow at most radii and as a result the disc will become  slim  \citep[rather than thin; e.g.,][]{Abramowicz88}. This could be the case with these two objects, specially in Mrk~142, where the measurements suggest an accretion rate close to the Eddington limit.
If this is the case, the disc temperature profile will change to $T(R)\propto R^{-1/2}$, and the time-lags spectrum becomes: $\tau(\lambda)\propto \lambda^2$ \citep[see][for a detailed discussion]{Cackett20}. 

The slope of 2 is quite different from the K20 model predictions (see Figure~\ref{fig:slope}). The fact that the slopes of the observed time-lags spectra in NGC~7469 and in Mrk~142 are fully consistent with our model suggests that the disc structure in these two objects is consistent with the standard, optically thick, geometrically thin disc model. In addition, the fact that the normalization is also consistent with the K20 models (for the given accretion rate) suggests that the X-ray source is located quite high above the disc, at a height of $\sim 50$ and $75$~\rg\ in Mrk~142 and NGC~7469, respectively.

\subsection{The BH spin} 

The K20 models fit the observed time-lags well both for $a^\ast=0$ and 1. On their own, the time-lags spectra do not favor either a static or a rotating BH. Figure~\ref{fig:mdot} shows that the $a^\ast=1$ best-fit $\dot{m}_{\rm Edd}$ values are always larger than the $a^\ast=0$ values. This is because we measure the accretion rate in Eddington units and have kept the BH mass fixed during the model fits. For a fixed $M_{\rm BH}$ and $\dot{m}_{\rm Edd}$, the accretion rate in physical units ($\rm M_\odot~yr^{-1}$) is smaller, and the outer disc is colder, in the case of a spinning BH. This is because the radiative efficiency increases with increasing spin. Consequently,  the $a^\ast=1$ accretion rates must have a larger value so that the disc temperature will be similar to the $a^\ast=0$ disc temperature which fits the data well. 

The degeneracy can be lifted when we compare the accretion rates with past estimates, which are usually based on the $L_{\rm Bol}/L_{\rm Edd}$ ratios. As we argued in Section~\ref{sec:bestfitres}, our estimates should not be significantly larger, or smaller than a factor of two (at most), than the $L_{\rm Bol}/L_{\rm Edd}$ based  estimates. Based on this comparison, our results imply a Schwarzschild (or a slowly rotating) BH in NGC~2617 and NGC~7469, and a rapidly rotating BH in Mrk~149, Mrk~509, NGC~5548 and NGC~4593 (4 out of 7 sources in the sample).  Even the $3\sigma$ upper limit is smaller than the accretion rate estimates which are based on the  $L_{\rm Bol}/L_{\rm Edd}$ ratios in these sources (see the horizontal dashed lines in Figure~\ref{fig:mcmc}).  

These results are suggestive at the moment. On the one hand, the error on \mdot\ is large (due to the large errors on the observed time-lags). It would be highly desirable to decrease the uncertainty on the observed time-lags, perhaps with the use of improved techniques, using the existing data sets. On the other hand, there is also a considerable uncertainty on the $L_{\rm Bol}$ estimates. They are usually based on luminosity measurements in a particular band, and the use of (quite uncertain) conversion factors. There are also some uncertainties in the computation of $L_{\rm Edd}$, because of the unavoidable errors on BH mass measurements. Nevertheless, our results suggest that modeling of the reverberation, UV/optical time-lags can be very helpful in measuring the BH spin in AGN. We plan to pursue this issue further in a future work.

\subsection {The X--ray corona height}

The uncertainty on the observed time-lags is rather large, so we cannot put strong constraints on the corona height. We can put somewhat stronger constrains on the source height in NGC~4593, because the time-lag errors are smaller for this source and the number of time-lags is large. We found that $h<47$~\rg\ and $h<62$~\rg\ ($3\sigma$ upper limit) in this case, for $a^\ast=0$ and 1, respectively. 
Nevertheless, the best-fit $h$ is larger than 10~\rg\ in {\it all} sources. This result indicates that, on average, the corona height is indeed large in these sources. This is opposite to what is observed in sources where X--ray reflection time-lags are detected, where the mean best-fit heights is equal or smaller than 5\rg\ \citep[e.g.,][]{Emmanoulopoulos14, Epitropakis16, Chainakun16, Caballero18}.  

NGC~5548 is the only object in our sample with X--ray reflection time-lags fit results. \cite{Emmanoulopoulos14} report a best-fit height of $\sim 5$~\rg, using data from an {\it XMM-Newton} observation that was taken in 2001. This is consistent with our results for spin 1  (at the $2\sigma$ level), although in this case, the inferred accretion rate would be quite larger than the value listed in Table~\ref{tab:sources}. The {\it XMM-Newton} observation was taken 13 years before the {\it Swift}, {\it HST} and ground based telescope observations that were used to compute the time-lags spectrum that we fit in this work. On the other hand, the X-ray spectral analysis of the \textit{Suzaku} monitoring of NGC~5548, presented by \cite{Brenneman12}, suggest a corona height larger than $\sim 100$~\rg. All this may be an indication that the X--ray source location varies with time in this source. \cite{Alston20} and \cite{Caballero-Garcia20} have shown that the corona height may change by a factor of $\sim 2-3$ in IRAS 13224$-$3809. In fact, \cite{Panagiotou20} detected a significant non-stationary behavior in the UV light curves of NGC~5548, and they also suggested the possibility of a variable corona height in this source.

The large height values we find are probably due to a selection effect. The objects in our sample show well defined UV/optical delays which are fully consistent thermal reverberation delays. This implies that the thermal reverberation signal is strong in these objects. According to K20, the amplitude of the variable, thermal reverberation component is expected to be significant in sources where the corona height is large. Hence the `bias' towards large corona heights for our sources.

\subsection {The X--ray vs UV/optical lags}

When extrapolated to the X--ray band, the model is not consistent with the observations. The observed X--ray/UV time-lags (the leftmost grey points in Figure~\ref{fig:lags}) are usually larger than what the K20 best-fit models predict. We believe that this is due to the fact that the cross-correlation function (CCF) between the X--ray and the UV/optical light curves depends on the corona/disc transfer function and on the X--ray auto-correlation function \citep[ACF; see eq. 7 in K20 and eq. 16 in][]{Peterson93}. The X-ray ACF is expected to be quite broad in AGN. The X--ray power-spectra in AGN have a power-law like shape, that extends to low frequencies (i.e., timescales of the order of years) with a slope of $\sim -1$ \citep[e.g.,][for NGC~4051]{McHardy04}.  This implies a very long memory in the X--ray emission process, and hence a very broad ACF. As a result, the X-ray ACF should significantly affect  the time-lags between X-rays and the UV/optical. However, the X-ray  ACF  should affect the CCF between X--rays and any UV/optical band in the same way. We  would  expect  the X-ray  ACF  to  cause  the same systematic shift to the time-lags between X--rays and any UV/optical band. It is for this reason that we believe the X--ray ACF does not affect the time-lags within the UV/optical bands, as the time-shifts due to the X--ray ACF should cancel out. Thus, it is preferable, when fitting the observed time lags using Equation~\ref{timelagseq}, to use a reference band that lies in the UV/optical range.

The NGC~7469 time-lags spectrum supports this explanation. The time-lags were estimated with respect to the X--rays in this source, but the long term X--ray variations were filtered out, using a 5-day filtering timescale \citep{Pahari20}. This filtering process should reduce the width of the X--ray ACF. Consequently, the K20 model fits very well the observed time-lags spectrum in this case, despite the fact that the X--rays are the reference band (bottom panel in Figure~\ref{fig:lags}). The X--rays/UV time-lags in NGC~5548 and NGC~2617 are also consistent with the best-fit K20 model, but this may be due to the large error of the respective time-lag (especially in the case of NGC~2617). 

\begin{figure}\centering
	\includegraphics[width=0.9\linewidth]{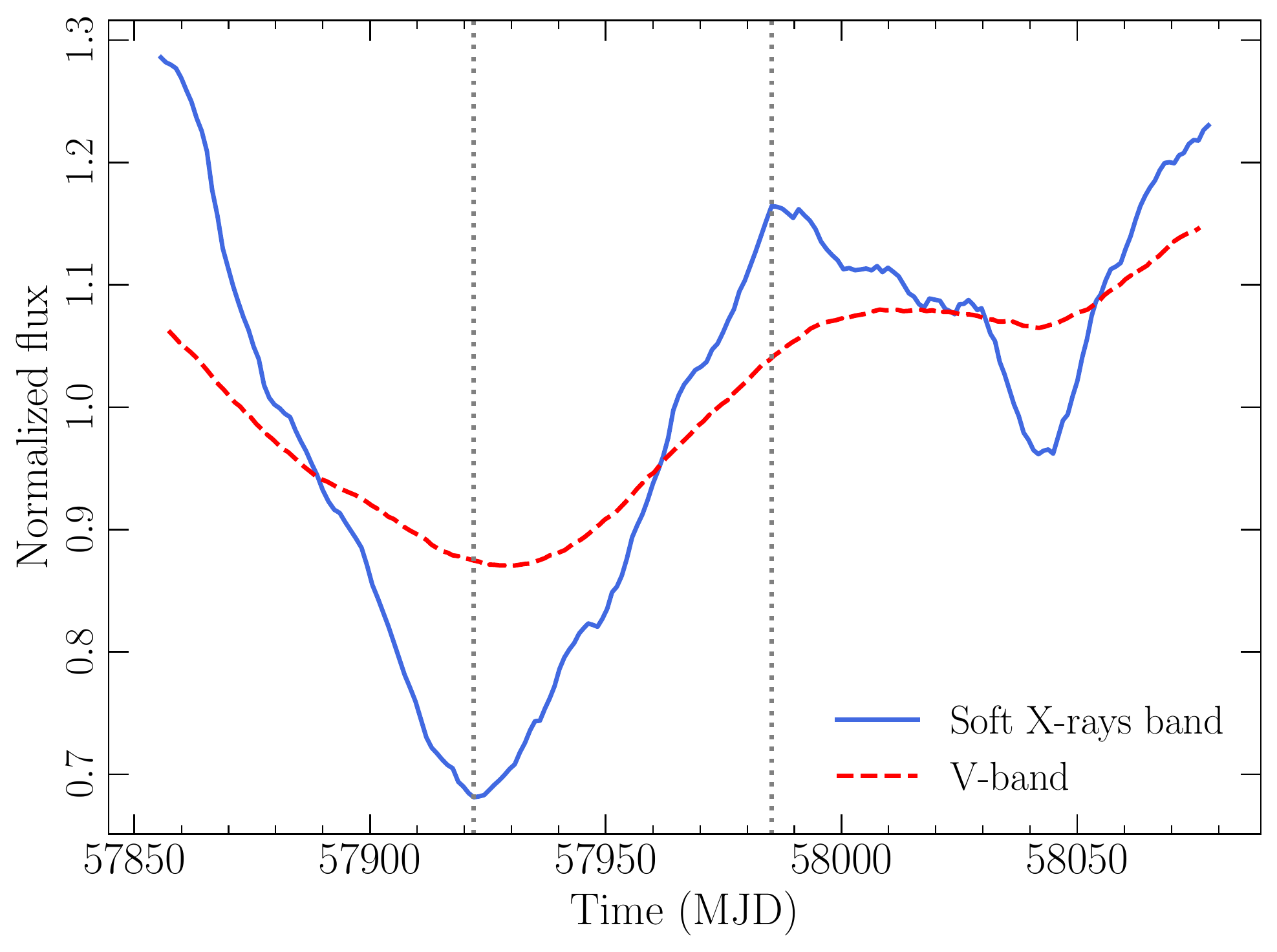}
    \caption{Mrk~509 smoothed light curves in the soft X-rays band (solid line) and the $V-$band (dashed line). The vertical lines indicate the first dip and the subsequent flux peak in the X–ray band.}
    \label{fig:mrklcs}
\end{figure}

The only exception is Mrk~509, where the X--rays appear to lag behind the UV variations (middle panel in the top row in Figure~\ref{fig:lags}). This cannot be explained by the K20 model (or any disc reverberation model). According to the E19 results, both the (soft band) X-rays and the $V-$band lag the UV variations by $\sim 2.5$~days. Figure~\ref{fig:mrklcs} shows the (soft band) X--ray and the V-band light curve of Mrk~509, normalized to the respective mean (data taken from E19). The light curves have been smoothed using a simple top-hat filter, 30-days wide. The vertical lines indicate the first dip and the subsequent flux peak in the X--ray band. We believe it is rather clear that the variations in the two bands are not simultaneous. The X-rays precede the respective features in the V-band light curve by a few days. A proper CCF analysis is out of the scope of this paper. Figure~\ref{fig:mrklcs} simply indicates that the X--ray/UV/optical correlation in  this source may be more complex than in other systems. We plan to revisit the X--ray/UV time-lags in this object in the future, to investigate in more detail the apparent discrepancy between the model and the observed X--ray time-lags. 

Another result from the CCF analysis of the monitoring observations we studied in this work is that, in general, the X-ray  to  UV  correlation  is  weaker  than  the UV to optical correlation. Arguably, the best way to address this issue is to investigate whether the observed X--ray light curves can indeed predict the shape (and amplitude) of the observed UV/optical variations. This is a challenging task, as the sampling rate of the observed light curves must be significantly denser than the width of the model transfer functions. Even the recent observations may not fulfil this requirement to address this issue properly (specially in the UV bands). We plan to investigate this issue in a future publication.

\section*{Acknowledgements}

EK acknowledges financial support from the Centre National d'Etudes Spatiales (CNES) from which part of this work was completed. MD thanks for the support from the GACR project 21-06825X and the institutional support from RVO:67985815. IEP acknowledges support  of  the  International  Space  Science  Institute,  Bern. This work made use of data supplied by the UK \textit{Swift} Science Data Centre at the University of Leicester. We would like to thank M. Pahari for providing the NGC~7469 time-lag data. The contour plots are plotted using the {\tt GetDist} Python package \citep{Lewis19}.

\section*{Data Availability}

The X-ray spectra analyzed in Section~\ref{sec:Xrayspec} can be all retrieved from the publicly available,  automatic \textit{Swift}/XRT data products generator \url{https://www.swift.ac.uk/user_objects/}. The data for the time-lag spectra are all published in previous works cited in Table~\ref{tab:sources}.


\bibliographystyle{mnras}
\bibliography{reference} 


\bsp	
\label{lastpage}
\end{document}